\documentclass[11pt,aps,nofootinbib,preprint,superscriptaddress]{revtex4}
\usepackage{color}
\usepackage{graphicx}
\usepackage{amsmath,amssymb,amsthm,amscd}
\newcommand{\IP}{\relax{\rm I\kern-.18em P}}
\newcommand{\IR}{\relax{\rm I\kern-.18em R}}

\def\be{\begin{equation}}
\def\ee{\end{equation}}

\def\bea{\begin{eqnarray}}
\def\eea{\end{eqnarray}}

\def\nn{\nonumber}
\def\ov{\over}
 
\def \ddel#1#2{\frac{\partial^2 #1}{\partial #2^2}}

\def\vev#1{\langle #1 \rangle}

 \usepackage{setspace} 

\newcommand{\CO}{{\cal O}}

\def\IR{{\mathbb R}}

\def\IP{{\mathbb P}}

\newcommand{\ri}{{\rm i}}

\newcommand{\tr}{{\rm Tr}}

\newcommand{\ba}{\begin{aligned}}
\newcommand{\ea}{\end{aligned}}

\newcommand{\ben}{\begin{eqnarray}\displaystyle}
\newcommand{\een}{\end{eqnarray}}

\begin{document}
\preprint{CERN-PH-TH/2006-078}
\preprint{TIFR/TH/06-01}

\title{Blackhole/String Transition for the Small Schwarzschild \\ Blackhole of $AdS_5 \times S^5$ and Critical Unitary Matrix Models}

\author{Luis \'Alvarez-Gaum\'e}
\email{Luis.Alvarez-Gaume@cern.ch}
\affiliation{CERN, CH-1211, Geneva, Switzerland}

\author{Pallab Basu}
\email{pallab@theory.tifr.res.in}
\affiliation{Tata Institute of Fundamental Research, Homi Bhaba Road, Mumbai-400005, India}

\author{Marcos Mari\~no}
\email{Marcos.Marino.Beiras@cern.ch}
\affiliation{CERN, CH-1211, Geneva, Switzerland}

\author{Spenta R. Wadia}
\email{wadia@theory.tifr.res.in}
\affiliation{Tata Institute of Fundamental Research, Homi Bhaba Road, Mumbai-400005, India}

\vspace*{5.0ex}

\begin{abstract}
In this paper we discuss the blackhole-string transition of the small
Schwarzschild blackhole of $AdS_5\times S^5$ using the AdS/CFT
correspondence at finite temperature. The finite temperature gauge
theory effective action, at weak {\it and} strong coupling, can be
expressed entirely in terms of constant Polyakov lines which are  $SU
(N)$ matrices. In showing this we have taken into account that there
are no Nambu-Goldstone modes associated with the fact that the 10 dimensional 
blackhole solution sits at a point in $S^5$. We show that the phase
of the gauge theory in which the eigenvalue spectrum has a gap
corresponds to supergravity saddle points in the bulk theory. We
identify the third order $N = \infty$ phase transition with the
blackhole-string transition. This singularity can be resolved using a
double scaling limit in the transition region where the large N
expansion is organized in terms of powers of $N^{-2/3}$. The $N =
\infty$ transition now becomes a smooth crossover in terms of a
renormalized string coupling constant, reflecting the physics of
large but finite N. Multiply wound Polyakov lines condense in the
crossover region. We also discuss the implications of our results for
the resolution of the singularity of the Lorenztian section of the
small Schwarzschild blackhole.

\end{abstract}

\maketitle

\section {Introduction and synopsis}
The problem of the fate of small Schwarzchild blackholes is important to understand, in a quantum theory of gravity. In a 
unitary theory this problem is the same as the formation of a small blackhole. An understanding of this phenomenon 
has bearing on the problem of spacelike singularities in quantum gravity and also (to some extent) on 
the information puzzle in blackhole physics. It would also teach us something about non-perturbative string physics.  

In the past Susskind \cite{Susskind:1993ws}, Horowitz and Polchinski (SHP) \cite{Horowitz:1996nw} and others 
\cite{Sen:1995in,Bowick:1986km,Bowick:1985af} have discussed this, in the framework of string theory, as a blackhole-string 
transition or more appropriately a crossover. Their proposal is that this crossover is parametrically smooth and it simply 
amounts to a change of description of the same quantum state in terms of degrees of freedom appropriate to the strength 
of the string coupling. The entropy and mass of the state change at most by $o(1)$. By matching the entropy formulas for 
blackholes and perturbative string states, they arrived at a crude estimate of the small but non-zero string coupling at the crossover. 
The SHP description is difficult to make more precise because a formulation of string theory in the crossover regime is not yet explicitly known.

There are many studies on the blackhole--string transition and the nature of the blackhole singularity in the case of two and three--dimensional blackholes\cite{Giveon:2005jv, Giveon:2003ge,Kutasov:2005rr,Giveon:2005mi,Nakayama:2004ge,Nakayama:2005pk}. 
Small extremal supersymmetric blackholes have been discussed in string theory with enormous success \cite{Dabholkar:2004yr,Dabholkar:2004dq, DDMP, Mohaupt:2005jd}. In particular the $\alpha'$ corrections to the entropy of the supersymmetric string sized blackholes has been matched to the microscopic counting.    

In this paper we discuss the blackhole--string crossover for the small 10 dimensional Schwarzschild blackhole in the 
framework of the AdS/CFT correspondence. In \cite{Alvarez-Gaume:2005fv}, building on the work of \cite{Sundborg:1999ue,Polyakov:2001af,Aharony:2005bq,Liu:2004vy, 
Spradlin:2004pp,Hallin:1998km}, a simplified model for the thermal history of small and big blackholes in $AdS_5$ 
(which were originally discussed by Hawking and Page \cite{Hawking:1982dh}) was discussed in detail . In particular, 
the large $N$ Gross-Witten-Wadia (GWW) transition \cite{Gross:1980he,Wadia:1979vk,Wadia:1980cp} was identified with the SHP transition for the small $AdS_5$ blackhole.

However it turns out that the small blackhole in $AdS_5$, which is uniformly spread over $S^5$, has a Gregory-Laflamme 
instability\footnote{This point was brought to our attention by O. Aharony and S. Minwalla. Understanding Gregory-Laflamme transition from a boundary perspective is an important open issue. In the current work we will not try to address this and only assume the existence of such a transition.}. When the horizon radius $r_h\sim 0.5R$ \cite{Hubeny:2002xn} the $l=1$ perturbation is unstable. The final configuration this instability leads to, as $r_h$ decreases and the horizon becomes less and less uniform over $S^5$, is most likely to be the 10 dim Schwarzschild blackhole. This small 10 dim Schwarzschild black hole does not have any further instability of Gregory-Laflamme type. This blackhole also happens to be a solution with asymptotic $AdS_5 \times S^5$ geometry for $l_s \ll r_h \ll R$(\ref{AdS}). 

When the horizon of this blackhole approaches the string scale $l_s$, we expect the supergravity (geometric) 
description to break down and be replaced by a description in terms of degrees of freedom more appropriate at 
this scale. Presently we have no idea how to discuss this crossover in the bulk IIB string theory. 
Hence we will discuss this transition and its smoothening in the framework of a general finite temperature 
effective action of the dual $SU(N)$ gauge theory on $S^3\times S^1$. In fact it is fair to say that in the crossover region we are 
really using the gauge theory as a definition of the non-perturbative string theory.

At large but finite $N$, since $S^3$ is compact, the partition function and all correlation functions are 
smooth functions of the temperature and other chemical potentials. There is no phase transition. 
However in order to make a connection with a theory of gravity, which has infinite number of degrees 
of freedom, we have to take the $N\rightarrow \infty$ limit and study the saddle point expansion in 
powers of $\frac{1}{N}$. It is this procedure that leads to non-analytic behavior. It turns out that by taking 
into account exact results in the $\frac{1}{N}$ expansion it is possible to resolve this singularity and 
recover a smooth crossover in a suitable double scaling limit. 

In the specific problem at hand, it turns out that in the transition region the large $N$ expansion is organized 
in powers of $N^{-2/3}$. In the bulk theory, assuming AdS/CFT, this would naively mean a string coupling 
expansion in powers of $g_s^{2/3}$. However in a double scaling limit, a renormalized string coupling 
$\tilde{g}=N^{\frac{2}{3}}(\beta_c-\beta)$ once again organizes the coupling constant expansion in integral powers. 
The free energy and correlators are smooth functions of $\tilde g$.

The use of the AdS/CFT correspondence for studying the blackhole-string crossover requires that there is a 
description of small Schwarzschild blackholes as solutions of type IIB string theory in $AdS_5 \times S^5$. 
Fortunately, Horowitz and Hubeny \cite{Horowitz:2000kx} have studied this problem with a positive conclusion. 
This result enables us to use the boundary gauge theory to address the crossover of the small Schwarzschild blackhole into a state described in terms of 'stringy' degrees of freedom. 
Even so the gauge theory is very hard to deal with as we have to solve it in the $\frac{1}{N}$ expansion for large but finite values of the 'tHooft coupling $\lambda$.
\par
However there is a window of opportunity to do some precise calculations because it can be shown that the 
effective action of the gauge theory at finite temperature can be expressed entirely in terms of the Polyakov loop 
which does not depend on points on $S^3$. This is a single $N\times N$ unitary matrix, albeit with a complicated interaction among the winding modes $\tr U^n$. This circumstance, that the order parameter $U$ in the gauge theory 
is a constant on $S^3$, matches well on the supergravity side with the fact that all the zero angular momentum 
blackhole solutions are also invariant under the $SO(4)$ symmetry of $S^{3}$. The blackhole may be localized in $S^5$, 
but it does not depend on the co-ordinates of $S^3$. The coefficients of the effective action depend upon the temperature, 
the `t Hooft coupling $\lambda$ and the vevs of the scalar fields. 
Since the 10-dimensional blackhole sits at a point in $S^5$, one may be concerned about the spontaneous breaking of 
$SO(6)$ R-symmetry and corresponding Nambu-Goldstone modes. We will conclude, using a supergravity analysis, that 
the symmetry is not spontaneously broken. Instead we have to introduce collective coordinates for treating the zero modes associated with this symmetry. 

The general unitary matrix model can be analyzed due to a technical progress we have made in 
discussing the general multi-trace unitary matrix model. We prove an identity that enables us to express 
and study critical properties of a general multi-trace unitary matrix model in terms of the critical properties of a general single trace matrix model.

As is well known, the single trace unitary matrix model at $N=\infty$ has a third order GWW transition, which occurs when the density of eigenvalues of the unitary matrix develop a gap on the unit circle. The vanishing of the density at a point on the circle leads to a relation among the coupling constants of the matrix model which defines a surface in the space of couplings (parameters of the effective action). The behavior of the matrix model in the neighborhood of this surface (call it critical surface) is characterized by universal properties which are entirely determined by 
the way the gap in the eigenvalue density opens: $\rho (\theta ) \sim (\pi -\theta) ^{2m}$, where $m$ is a 
positive integer. In our problem, there is only one tunable parameter, namely the temperature. 
Hence we will mainly focus only on the lowest $m=1$ critical point and present the relevant operator that opens the gap. 
We also discuss the possible relevance of higher order multi-critical points.

Using the properties of the $\frac{1}{N}$ expansion near and away from the critical surface, we will 
argue that the small blackhole (or for that matter any saddle point of supergravity around which a well 
defined closed string perturbation expansion exists) corresponds to the phase of the matrix model where 
the density of eigenvalues on the unit circle has a gap. The small blackhole therefore corresponds to the gapped phase of the unitary matrix model.

We make a reasonable physical assumption based on the proposal of SHP, that the thermal history of the unstable saddle point corresponding to the small blackhole, eventually intersects the critical surface at a critical temperature $T_c$, which is $o(1/l_s)$. $T_{c}$ is smaller than the Hagedorn temperature. Once the thermal 
history crosses the critical surface it would eventually meet the $AdS_5\times S^5$ critical point corresponding to a uniform eigenvalue distribution.
(Such a history was already discussed in the context of a simplified model in \cite {Alvarez-Gaume:2005fv}.) It is natural to identify the crossover 
across the critical surface in the gauge theory as the bulk blackhole-string crossover. 


At the crossover, the $o(1)$ part of the gauge theory partition function (which depends on the renormalized string coupling) can be exactly calculated in a double scaling limit. This is a universal result in a sense that it does not depend on the location of the critical point on the critical surface but depends only on deviations which are normal to the critical surface. If we parametrize this by $t$, the the free energy $-F(t)$ solves the differential equation $\ddel{F}{t} = -f^2(t)$ where $f(t)$ satisfies the Painlev\'e II equation. The exact analytic form of $F(t)$ is not known, but $F(t)$ is a smooth function in the domain $(-\infty,\infty)$\footnote{This universal formula also appeared in the discussion of the simplified model in \cite {Alvarez-Gaume:2005fv}}. All the operators $\rho_k=\frac{TrU^{k}}{N}$ condense in the crossover region. In fact $\vev{N^{\frac{2}{3}}(\rho_k-\rho^{ug}_k)}=C_k \frac{d}{dt}F$, where  $C_k=\frac{(-1)^k}{k}$ and $\rho^{ug}_k$ represents the expectation value of $\rho_k$ in the ungapped phase.

The smooth crossover of the Euclidean blackhole possibly has implications for the resolution of the singularity of the Lorentzian blackhole, because within the $AdS/CFT$ correspondence we should be able to address all physical questions of the bulk theory in the corresponding gauge theory. In particular we should be able to address phenomena both outside and inside the blackhole horizon.

The plan of this paper is as follows. Section 2 discusses the SHP transition. Section 3 discusses the small 10-dimensional blackhole in $AdS_5\times S^5$. Section 4 discusses the finite temperature gauge theory and the effective action in terms of the unitary matrix model. Section 5 presents the multi-trace partition function as the calculable integral transform of the single trace unitary matrix model.
Section 6 discusses critical behavior in the unitary matrix model. Section 7 discusses the saddle point equations of the matrix model. Section 8 discusses the double scaled partition function. Section 9 discusses the introduction of chemical potentials and higher critical points. Section 10 discusses the applications of the critical matrix model to the small 10-dimensional blackhole. Section 11 discusses the Lorentzian blackhole.

\section{Blackhole-string transition}\label{sec:StBh}
\
\par
In this section we review the blackhole-string crossover. Consider the 10-dim Schwarzschild blackhole. As long as its horizon radius $r_h \gg l_s$  ($l_s$ is the string length), 
the supergravity description is valid and we can trust the lowest order effective action in $l_s$. When $r_h  \sim l_s $, this description breaks down and 
one learns to derive an effective action valid to all orders in $l_s $ or devises other methods to deal with the problem. Let us assume that the all orders in $l_s$ description is available, 
then presumably the geometrical description is still valid in principle, and one can indeed discuss the notion of a string size horizon with radius $
r_h \sim l_s$ \cite{Dabholkar:2004yr,Dabholkar:2004dq,Mohaupt:2005jd}. 
It is reasonable to expect that in such a description the qualitative fact that the mass decreases with the horizon radius and increasing temperature, is still valid. 
These facts are obviously valid to lowest order in $l_s$, because $r_h = 2G_{N}M$ and $T_h = (G_NM)^{1/7}$. Here $G_N$ is Newton's coupling and $M$ is 
the mass of the blackhole. For definitiveness let us fix the mass and the entropy of the blackhole. Then the $r_h$ and $T_h$ vary with the gravitational coupling $G_N$. 
Now since $g_s^2 = G_N l_s^{-8}$, we can say that  $r_h$ and $T_h$ vary with $g_s$ and hence a crossover at $r_h\sim l_s$ happens at a specific value of the string coupling.

When $r_h \lesssim l_s$ the above description of the state has to be replaced by a description in terms of microscopic degrees of freedom relevant to the scale $l_s$. Even in this description it is reasonable to assume that the temperature of the state varies as we change the string coupling. The assumption of Susskind-Horowitz-Polchinski is that the mass of the state would change by at most o(1) in the string coupling.


From the above discussion it is clear that the blackhole-string crossover occurs in a regime where the curvature of the blackhole 
is $o(1)$ in string units, so as to render the supergravity description invalid. It is also clear that besides $l_s$ related effects, the string coupling is non-zero and its effects have to be taken into account. 
Presently our understanding of string theory is not good enough for us to make a precise and quantitative discussion of the crossover. Hence we will discuss the problem using the AdS/CFT 
correspondence. In order to do this we need to be able to embed the small blackhole in $AdS_5\times S^5$. This has been discussed by Horowitz and Hubeny \cite{Horowitz:2000kx}. 
We briefly review their work in the next section.

\section{Embedding the 10-dimensional Schwarzschild blackhole in $AdS_5\times S^5$}
It is not difficult to argue that the small 10-dim Schwarzschild blackhole exists as a solution of Einstein's equation in $AdS_5 \times S^5$. A small patch of the $AdS_5 \times S^5$ space is locally identical to 10 dim Euclidean space. Since the horizon radius of this blackhole $r_h \ll R$, we can have a solution which is locally identical to a 10 dim Schwarzschild blackhole in flat space-time. We would also require that the solution for large 10 dimensional radial distances asymptotes to $AdS^5 \times S^5$. This solution is not explicitly known, but can be numerically constructed given the boundary conditions on the radial functions. The more non-trivial issue is concerning the fact that the type IIB theory also has a 5-form. In the absence of the blackhole this form is the volume form of $S^5$ and carries N units of flux. It turns out that in the presence of the small blackhole, a consistent solution to the equations of motion, is such that there is no energy flux into the blackhole. Hence the small blackhole remains small. In the above analysis one neglects the back reaction on the metric due to the fact that the blackhole is small and the curvature near its horizon is large.

The solution is conveniently represented if we use a $10$ dimensional radial coordinate system (fixed by the area of $S^8$) in $AdS_5 \times S^5$. One splits $S^8$ into $S^3$ and $S^4$, corresponding to the rotational $SO(4)$ symmetry of $AdS_5$ and the remaining (unbroken) $SO(5)$ symmetry of $S^5$. This is achieved by using the following coordinate transformation in (\ref{AdS}) 
\bea
r &=& \rho \sin\theta \\
\nn \chi &=& \rho \cos\theta
\eea
In these coordinates, a flat patch within $AdS$ is achieved in the limit $r,\xi \ll R$, where R is the radius of  $AdS_5$. The metric takes the form 
\bea
ds^2=-dt^2+d\rho^2+\rho^2(d\theta^2+\sin^2\theta d\Omega_3^2+\cos^2\theta d\Omega_4^2)
\eea
(The angular term in parenthesis is equivalent to $d\Omega_8^2$). Similarly the 5-form field strength takes the form 
\bea
F=-\rho^3\sin^4\theta dt \wedge d\rho \wedge d\Omega_3-\rho^4\sin^3\theta\cos\theta dt \wedge d\theta \wedge d\Omega_3+\\
\nn r^4\cos^5\theta d\rho \wedge d \Omega_4 - r^5 \sin\theta \cos^{4}(\theta) d\theta \wedge d \Omega_4
\eea
In this metric the Schwarzschild solution is given by 
\bea
ds^2=-f(\rho)dt^2+f^{-1}(\rho)d\rho^2+\rho^2(d\theta^2+\sin^2\theta d\Omega_3^2+\cos^2\theta d\Omega_4^2) \\
F= g_1(\rho,\theta)[-\rho^3\sin^4\theta dt \wedge d\rho \wedge d\Omega_3 - r^5 \sin\theta \cos^{4}\theta d\theta \wedge d \Omega_4] \\ 
\nn + g_2(\rho,\theta)[\rho^4\sin^3\theta\cos\theta dt \wedge d\theta \wedge d\Omega_3+r^4\cos^5\theta d\rho \wedge d \Omega_4 ]
\eea
where near the blackhole horizon $f=1-\frac{r_{h}^7}{r^7}$. As $r \rightarrow \infty$, the functions $f(r),g_1(r,\theta),g_2(r,\theta)$ approach their corresponding values in $AdS_5 \times S^5$ geometry. The explicit solution for these functions are not known but their form can be determined by numerically integrating a set of coupled linear differential equations. These solutions have the desired property that imply that the small blackhole remains small.

\section{Finite temperature gauge theory, order parameter and effective action}

We first present a general discussion of the order parameter of $SU(N)$ YM theory on the compact manifold $S^3$. We consider the theory in the canonical ensemble, i.e. the Euclidean time direction is periodically identified with a period of $\beta = {1 \ov T}$. It was pointed out in \cite{Sundborg:1999ue,Aharony:2003sx} that the Yang-Mills theory partition function on $S^3$ at a temperature $T$ can be reduced to an integral over a unitary $SU(N)$ matrix $U$, which is the zero mode of Polyakov loop on the euclidean time circle. Their analysis was done in the limit when the 'tHooft coupling $\lambda \rightarrow 0$.
 \bea
  Z (\lambda,T)= \int dU \, e^{ S (U)}
  \eea
with
 \bea
 U = P \exp \Bigl(i \int_0^{ \beta} A_{0} d \tau \Bigr)
 \eea
where $A_{0} (\tau) $ is the zero mode of the time component of the gauge field on $S^3$. This follows from the fact that apart from $A_{0}$ all modes of the gauge theory on $S^3$ are massive. We will discuss the validity of the above expression in both strong and weak ($\lambda$) coupling regimes. Hence we can use $U$ as an order parameter. 
Gauge invariance requires that the effective action of $U$ be expressed in terms of products of $\tr U^n$, with $n$ an integer, since these
are the only gauge invariant quantities that can be constructed from
$A_{0}$ alone. $S_{\rm eff} (U)$ also has a $Z_N$ symmetry under
\bea
U \to e^{{2 \pi i \ov N}} U \
\eea
This is due to the global gauge transformations which are periodic in the 
Euclidean time direction up to $Z_N$ factors. $Z_{N}$ invariance puts further restriction on the form of the effective action and a generic term in
$S(U,U^{\dagger})$ will have the form
\bea
\tr\, U^{n_1} \tr \, U^{n_2} \cdots \tr U^{n_k}, \qquad n_1 + \cdots
n_k = 0 \; ({\rm mod} \; N), \qquad k>1
\eea
In the large N limit we can work with $U(N)$ rather than $SU(N)$, and in that case $Z_N$ is replaced by $U(1)$.

We can expand $S_{\rm eff}$ in terms of a complete set of such operators. The first few terms are
 
\bea
\nn S(U,U^{\dagger})=a\tr \,U \tr \, U^{-1}+\frac{b}{N}\tr \, U^{2}\tr \, U^{-1}\tr \, U^{-1}+ \\
\frac{c}{N^2} \tr \, U^{3} \tr \, U^{-1} \tr \, U^{-1} \tr \, U^{-1}+\cdots
\label{EffAction}
\eea

More generally we will write the effective action (\ref{EffAction}) in a form which will be convenient for later discussion,
\bea
S (U,U^{\dagger}) = \sum_{i=1}^p a_i \tr \, U^i  \tr \, U^{\dagger i} + \sum_{\vec k, \vec k'} \alpha_{\vec k , \vec k'} \Upsilon_{\vec k} (U) \Upsilon_{\vec k'}(U^{\dagger}),
\label{genaction}
\eea
where $\vec k$, $\vec k'$ are arbitrary vectors of nonnegative entries, and
\be
\label{upsop}
\Upsilon_{\vec k} (U)=\prod_j \Bigl( \tr \, U^j \Bigr)^{k_j}.
\ee
It is useful to define
\be
\ell(\vec k) =\sum_j j k_j, \quad |\vec k| =\sum_j k_j.
\ee
The above parametrization of the general action is slightly redundant, since the second summand in (\ref{genaction}) is already the most general gauge-invariant 
action for $U$, $U^{\dagger}$, but writing it this way will be very useful. Reality of the action (\ref{genaction}) requires $\alpha_{\vec k \vec k'} =\alpha^*_{\vec k' \vec k}$. In fact, using the explicit perturbative rules to compute $S(U,U^{\dagger})$ in (\ref{genaction}), one can show that the $\alpha_{\vec k \vec k'}$ are real, therefore 
\be
\label{alphacond}
\alpha_{\vec k \vec k'} =\alpha_{\vec k' \vec k}.
\ee
On the other 
hand, invariance of $S (U,U^{\dagger})$ under $U\rightarrow e^{i \theta} U$ requires that
\be
\ell(\vec k) =\ell(\vec k'). 
\ee

We now present evidence at both weak and strong $\lambda$ that the above effective action is correct.
\par

\subsection{Perturbative analysis}
In perturbation theory one can integrate out all fields, except the zero-mode $A_0$ of the time component of a gauge field, to get an effective action of $U$ \cite{Aharony:2003sx}. All fields other than this mode are massive in a free YM theory on $S^{3}$. The scalar fields get their mass due to the curvature coupling. We can expand all other fields on $S^{3}$, and due to the finite radius of $S^{3}$ all the harmonics are massive. Hence at small coupling (small $\lambda$) one may integrate out all the fields and derive an effective action for $U$. In \cite{Aharony:2005bq} the perturbative (up to three loop order) effective action was calculated and it has the form (\ref{EffAction}).
\subsection{Strong coupling analysis}
The above discussion is perturbative and there is no guarantee that the scalar fields remain massive in the expansion of the theory around $\lambda=\infty$. 
We will now show, using the $AdS/CFT$ correspondence, that even at strong coupling (large $\lambda$), all the excitations of N=4 SYM theory on $S^3$ 
are massive \cite{Witten:1998zw}. For illustration we consider the wave equation of a scalar field $\phi(r,t)$ in a general blackhole background which is asymptotically $AdS_5 \times S^5$. 

The $AdS_5 \times S^5$ metric is given by
\bea
ds^2= (1+\frac{r^2}{R^2})d\tau^2+\frac{dr^2}{1+\frac{r^2}{R^2}}+r^2 d\Omega_3 ^2 +R^2 d\Omega_5^2
\label{AdS}
\eea

Let us consider the situation when the asymptotic solution depends on the co-ordinates of $S^5$ and $S^3$. Since $S^5$ and $S^3$ are compact spaces, their laplacians have a discrete spectrum. 
We focus on the radial part and consider a finite energy solution of energy $E$, $\phi(r,\theta_3,\theta_5,\tau)=f(r,\theta_3,\theta_5)\exp(E\tau)$. 
The wave equation in the asymptotic metric (\ref{AdS}) is given by
\bea
\nn (3+5r^2) f'(r,\theta_3,\theta_5)+r(1+r^2)f''(r,\theta_3,\theta_5)+(\frac{r}{1+r^2}E^2 +\frac{1}{r}\Delta_{\Omega3}^2+\\
 \nn r\Delta_{\Omega_5}^2)f(r,\theta_3,\theta_5)&=&0 \\
\nn (3+5r^2) f'(r)+r(1+r^2)f''(r)+(\frac{r}{1+r^2}E^2 -\frac{1}{r}L_{\Omega3}^2-r M_{\Omega_5}^2)f(r)&=&0 
\label{laplac}
\eea    
where $'$ is the partial derivative with respect to $r$ and $L_{\Omega3}$ is the contribution from $S^3$ harmonics and $M^2_{\Omega5}$ is the contribution from $S^5$ harmonic.   

For $f(r) \sim r^\alpha$, as $r \rightarrow \infty$, equation (\ref{laplac}) reduces to 
\bea
5r^{\alpha+2}( (\alpha (\alpha-1)+5\alpha)-M^2_{\Omega5})=0 
\label{aleq}
\eea
In the last equation we have neglected the term $E^2 r^\alpha$ and the $S^3$ harmonics part, as it is suppressed by a factor of order $\frac{1}{r}$. 
Hence $\alpha_{1}=-2+\sqrt{4+M^2}$ or $\alpha_{2}=-2-\sqrt{4+M^2}$ are two solutions of (\ref{aleq}). Consequently , $f(r) \sim r^{\alpha_2}$ is the only solutions which is normalizable. 
 
Let us now analyze the situation near the blackhole horizon which, in the euclidean continuation, acts like the origin of polar co-ordinates. Hence, we have the boundary condition,
\bea
\frac{df}{dr}=0
\label{nhr}
\eea   
Near the origin, the scalar field laplacian in the blackhole back ground will have two solutions for a given $E$. One of them diverges at the horizon and other maintains the condition (\ref{nhr}). For a generic $E$, a well-behaved solution in general approaches a non-renormalizable solution as $r \rightarrow \infty$. As in quantum mechanical problems, a normalizable solution exists only for those values of $E$ for which, the solution that behaves correctly at the lower endpoint also vanishes for $r \rightarrow \infty$. This eigenvalue condition determines a discrete value of $E$. Hence the mass gap in SYM theory on $S^3$ persists at the strong coupling. 
The basic physical reason for the discrete spectrum is that the asymptotic $AdS_5 \times S^5$ geometry gives rise to an infinitely rising potential for large $r$. 

In order to make an estimate of the mass gap we note that the blackhole metric depends on the combination $GM$, where $G\sim \frac{1}{N^2}$ is Newton's coupling and $M\sim N^2$ is the mass of the blackhole. Further using standard formulas of blackhole thermodynamics it is possible to express $GM$ entirely in terms of the temperature of the blackhole, which sets the scale of the mass gap.

We also expect the single negative eigenvalue in the spectrum of the euclidean Schwarzschild solution in asymptotically flat space-time to persist in the present case. Next we discuss the zero modes.

\subsubsection{$SO(6)$ non-invariance of the 10-dimensional blackhole}
 As discussed in the introduction, our main interest is the study of the 10 dimensional small blackhole to string transition in $AdS_5 \times S^5$. The metric of the 
 small 10 dimensional blackhole in $AdS_5 \times S^5$ is not symmetric under the $SO(6)$ transformations of $S^5$. Hence the corresponding 
 saddle point in the gauge theory would transform under the $SO(6)$ R-symmetry group and a natural question is whether the $SO(6)$ symmetry 
 is spontaneously broken in the dual gauge theory with associated massless Nambu-Goldstone modes. If this were true, then we would have to include 
 additional degrees of freedom in the effective action (\ref{EffAction}).
\par
Fortunately even though the small 10 dimensional blackhole sits at a point in $S^5$ the massless modes associated with motions about this point 
correspond to normalizable solutions of the small fluctuations equation. Let us discuss this point in more detail.

We have already discussed in the section 3 that the small 10 dim blackhole is invariant under an ``unbroken'' $SO(5)$ subgroup of $SO(6)$. The remaining broken generators of $SO(6)$ rotate the blackhole in $S^5$. The blackhole is labeled by its mass (equivalently temperature) and its position in $S^5$, which we denote by the co-ordinates $\theta_5$. $SO(6)$ rotations can rotate the blackhole to any point in $S^5$. The action of the initial and final blackhole is the same, because we get the final solution just by a co-ordinate rotation of the initial solution. As there is an orbit of blackhole solutions with the same action, it is expected that there is a zero mode in the spectrum of the small oscillations operator around the blackhole. 

Let us clarify this point in more detail. Consider a blackhole metric ($g^{0}_{\mu \nu}(\theta_5)$) as a function of $\theta_5$. As we mentioned before, an infinitesimal rotation in $S^5$ creates a new black solution which is given by $g^{1}_{\mu \nu}=g^{0}_{\mu \nu}+\delta g_{\mu \nu}$. As both the matrices $g^{0}_{\mu \nu}$ and $g^{1}_{\mu \nu}$ solve the equations of motion, their difference $\delta g_{\mu \nu}$ will be a zero mode. The existence of such a zero mode does not necessarily signal the onset of  spontaneous symmetry breaking. The important point is whether the zero mode is normalizable or not. We will show that $\delta g_{\mu \nu}$ is a {\it  normalizable} zero mode. 

\par
We make the assumption that the asymptotic geometry of an uncharged blackhole solution is determined by its mass. Hence the asymptotic geometry of the blackhole is given by that of a small $AdS_5$ blackhole \cite{Hawking:1982dh} with corrections $f_{\mu \nu}$,
\bea
ds^2 &=& (1+\frac{r^2}{R^2}-\frac{m}{r^2})dt^2+\frac{dr^2}{(1+\frac{r^2}{R^2}-\frac{m}{r^2})}+r^2 d\theta_3 ^2+R^2 d\theta_5^2 + f_{\mu \nu}dx^{\mu}dx^{\nu}
\eea    
where $f_{\mu \nu} \sim \frac{1}{r^3}$ as $r \rightarrow \infty$. Hence the difference of $g^{0}(\mu,\nu)$ and $g^{1}(\mu,\nu)$ can be written as 
\bea
\delta g(\mu,\nu)=f^{1}_{\mu,\nu}-f^{0}_{\mu,\nu}
\eea
where $f^{0}$ and $f^{1}$ denotes the $f$'s corresponding to $g^0$ and $g^1$. Now $f_{\mu \nu} \sim \frac{1}{r^3}$ implies $\delta g_{\mu \nu} \sim \frac{1}{r^3}$. Hence $\delta g_{\mu \nu}$ is square integrable \footnote{This argument seems to be independent of $\alpha'$ corrections as the asymptotic geometry is always weakly curved for any black hole situated in a asymptotic $AdS$ space with $l_s << R$.},
\bea
\int d^4 x \delta g_{\mu \nu} ^2 \propto \int dr r^3 \frac{1}{r^6} \propto \int dr \frac{1}{r^3}
\eea 

Since the symmetry is not spontaneously broken, we should consider the full orbit of the classical field under SO(6) (or its coset) using the method of collective coordinates \cite{Gervais:1974dc}. Hence we have the situation in which the degrees of freedom correspond to two sets of zero modes: those corresponding to $A_0$ and those corresponding to $SO(6)$ symmetry. In the method of collective coordinates we make the following change of variables in the gauge theory path integral. 

For simplicity of presentation we denote the fields of the gauge theory that transform under $SO(6)$ by $\phi (x)$ and consider 
\bea
\phi(x)=\phi_{0}(x)^{[\Omega^5]}+\eta(x) \\
\eea
and the gauge condition,
\bea
(\eta, \phi_{0}^{[\Omega^5]})=0
\eea
where $\phi_{0}(x)^{[\Omega^5]}$ is the orbit under $SO(6)$ of the classical configuration $\phi_{0}(x)$. The path integral measure now becomes 
\bea
D\phi(x)=d\Omega^5 D\eta (x)\delta (\eta, \phi_{0}^{[\Omega^5]})\Delta
\eea
where $\Delta$ is the Faddev-Popov determinant. Then by standard means we can see that the zero mode is eliminated by the delta-function and the collective coordinate (compact group measure) factors out of the path integral and the remaining action is a functional of the classical field $\phi_{0}(x)$. Integrating out the fluctuations $\eta$, we will obtain an effective action entirely in terms of the unitary matrix $U$. The coefficients of the effective action will now depend on the vevs of the scalar fields. 

\subsection{Comments on the effective theory}
It should be mentioned that the effective action (\ref{EffAction}) is constructed only from the zero mode of $A_{0}$ on a compact manifold. Hence this effective action will not be able to describe physical situations which depend on the co-ordinates of the compact manifold $S^3$. However on the supergravity side all the zero angular momentum blackhole solutions are invariant under the $SO(4)$ symmetry of $S^{3}$. The blackhole may be localized in $S^5$, but it does not depend on the co-ordinates of $S^3$. This fortunate circumstance enables us to use (\ref{EffAction}) as a reliable effective action to describe some aspects of the string theory in $AdS_5 \times S^5$.
\par 
The saddle points of (\ref{EffAction}) corresponding to the N=4 SYM theory are in one to one correspondence with the bulk supergravity (more precisely IIB string theory) saddle points. For example, the $AdS_5 \times S^5$ geometry corresponds to a saddle point such that $\langle \tr \, U^n\rangle=0$ $\forall n \neq 0$. Hence the eigenvalue density function is a uniform function on the circle. Now, depending on the co-efficients in (\ref{EffAction}) the saddle point $\vev{\tr \, U^n}$ can have a non-uniform gaped or ungapped eigenvalue density profile. Changing the values of the coefficients, by varying the temperature, may open or close the gap and lead to non-analytic behavior in the temperature dependence of the free energy at $N=\infty$. We will interpret this phenomenon as the string-blackhole transition. As we shall see this non-analytic behavior can be smoothened out by a double scaling technique in the vicinity of the phase transition.

\section{Exact integral transform for the partition function}\label{exacttrans}

We start with the most general effective action given in the equation (\ref{genaction}). The partition function is given by 
\be
\label{genz}
Z=\int [d\, U] e^{S(U,U^{\dagger})}.
\ee
We will assume in the following that $a_i>0$ in (\ref{genaction}). This amounts to the assumption that $\rho_{i}=\langle \frac{1}{N}\tr \, U^i \rangle=0$ is always a saddle point of the effective action. It corresponds to the $AdS_5 \times S^5$ saddle point of IIB string theory. In \cite{Basu:2005pj} it was shown that, at sufficiently low temperatures, $a_{1}>0$.
\par
 We now use the standard Gaussian trick to write, 
\be 
\label{gauss1}
\ba
&\exp\biggl\{ \sum_{i=1}^p a_i {\rm tr}\, U^i {\rm tr}\, U^{\dagger i} \biggr\}  \\
&=  \biggl( {N^2\over 2\pi}\biggr)^p\int   \prod_{i=1}^p {dg_i \, d\bar g_i\over a_i} \exp\biggl\{ 
-N^2\sum_{i=1}^p {g_i \bar g_i \over a_i} + N \sum_{i=1}^p ( g_i {\rm tr}\, U^i +  \bar g_i {\rm tr}\, U^{\dagger i} ) \biggr\}
\ea
\ee
Using this trick a second time we have,
\bea
 \nn & & \exp(-N^2 \sum_{j=1}^p {g_j \bar g_j \over a_j })   \\
&=& \biggl( {N^2\over \pi}\biggr)^p\int   \prod_{j=1}^p a_j d\mu_j\, d\bar \mu_j \exp\biggl\{ -N^2 \sum_{j=1}^p a_j\mu_j  \bar \mu_j  + {\rm i} N^2\sum_j (\mu_j \bar g_j + \bar \mu_j g_j)\biggr\}
\label{gauss2}
\eea
In order to deal with an arbitrary polynomial $P$ of $\tr\, U^i, \tr \, U^{\dagger i}$, we use the following identity in (\ref{gauss1}),
\bea
 & & \nn \exp \biggl\{ P( \tr\, U^i ,\tr\, U^{\dagger i}) +\sum_{i=1}^p a_i {\rm Tr}\, U^i {\rm Tr}\, U^{\dagger i} \biggr\}   \\ 
&=& \biggl( {N^2\over 2\pi}\biggr)^p\int   \prod_{i=1}^p {dg_i \, d\bar g_i\over a_i} \exp\biggl\{-N^2\sum_{i=1}^p {g_i \bar g_i \over a_i}\biggr\} \exp\biggl\{P(\frac{\partial}{N\partial g_i},\frac{\partial}{N\partial \bar g_i}) \biggr\}\\ 
\nn & & \cdot \exp{\biggl\{ N \sum_{i=1}^p ( g_i {\rm Tr}\, U^i +  \bar g_i {\rm Tr}\, U^{\dagger i} )\biggr\}}\\
\nn &=& \biggl( {N^2\over 2\pi}\biggr)^p\int   \prod_{i=1}^p {dg_i \, d\bar g_i\over a_i} \exp\biggl\{ N \sum_{i=1}^p ( g_i {\rm Tr}\, U^i +  \bar g_i {\rm Tr}\, U^{\dagger i} )\biggr\} \\
\nn & &\cdot \exp\biggl\{P(-\frac{\partial}{N\partial g_i},-\frac{\partial}{N\partial \bar g_i}) \biggr\} \exp\biggl\{-N^2\sum_{i=1}^p {g_i \bar g_i \over a_i}\biggr\} \\
\eea
In the last line we have integrated by parts. Then we use (\ref{gauss2}) to write  
\bea 
& & \exp\biggl\{P(-\frac{\partial}{N\partial g_i},-\frac{\partial}{N\partial \bar g_i}) \biggr\} \exp\biggl\{-N^2\sum_{i=1}^p {g_i \bar g_i \over a_i}\biggr\} \\ 
\nn &=&  \biggl( {N^2\over \pi}\biggr)^p \exp\biggl\{P(-\frac{\partial}{N\partial g_i},-\frac{\partial}{N\partial \bar g_i})\biggr\}\\
& &\nn \cdot \int   \prod_{j=1}^p a_j d\mu_j\, d\bar \mu_j \exp\biggl\{ -N^2 \sum_{j=1}^p a_j\mu_j  \bar \mu_j  + {\rm i} N^2\sum_j (\mu_j \bar g_j + \bar \mu_j g_j)\biggr\} \\
\nn &=& \biggl( {N^2\over \pi}\biggr)^p\int   \prod_{j=1}^p a_j d\mu_j\, d\bar \mu_j \exp\biggl\{ -N^2 \sum_{j=1}^p a_j\mu_j  \bar \mu_j  + {\rm i} N^2\sum_j (\mu_j \bar g_j + \bar \mu_j g_j+P(\ri N\mu_j,\ri N\bar\mu_j)\biggr\}
\eea

Since the effective action (\ref{genaction}) is a polynomial in $\tr\, U^{i}$, $\tr\, U^{\dagger i}$, we can use the procedure discussed above to write the partition function (\ref{genz}) as
\bea
\label{fullint}
Z=\biggl( {N^4\over 2\pi^2}\biggr)^{p} \int   \prod_{i=1}^p dg_i \, d\bar g_i  \, d\mu_i \, d\bar \mu_i \exp (N^2 S_{\rm eff})
\eea
where
\bea
\label{effaction}
S_{\rm eff}&=&-\sum_{j=1}^p a_j \mu_j \bar \mu_j +{\rm i}\sum_j (\mu_j \bar g_j + \bar \mu_j g_j) + \\
\nn & & \sum_{\vec k, \vec k'}  \alpha_{\vec k , \vec k'} (-{\rm i})^{|\vec k| + |\vec k'|} \Upsilon_{\vec k} (\bar \mu) \Upsilon_{\vec k'}(\mu) + F(g_k , \bar g_k ).
\eea
In the above formula we have introduced the definition
\be
\Upsilon_{\vec k}(\mu)=\prod_j \mu_j^{k_j}.
\ee
and the free energy $F(g_k, \bar g_k)$ is defined by
\bea
\label{standardmodel}
\exp (N^2 F(g_k , \bar g_k)) &=& \int [d\, U] \exp\biggl\{ N \sum_{i\ge 1 } ( g_i  {\tr}\, U^i +  \bar g_i {\tr}\, U^{\dagger i} ) \biggr\},
\eea
It is important to note that given the effective action $S(U,U^{\dagger})$ of the gauge theory, $S_{\rm eff}$ can be exactly calculated. 

One notes that  $F(g_i, \bar g_i)$ depends only on the $p-1$ phases, since one of the phases of the $g_i$ can be absorbed by a rotation of $U$ in the unitary integral in (\ref{standardmodel}). 
The full integrand (\ref{fullint}) can be shown to be independent of one phase of $g_i$ by a redefinition of the auxiliary variables $\mu_j, \bar \mu_j$. 
\par 
The significance of (\ref{fullint}) is that the partition function (\ref{genz}) can be expressed as an exact integral transformation of the linear matrix model (\ref{standardmodel}). The phase structure and the critical behavior of the linear matrix model is well understood, and hence we can study these to learn about the critical behavior and the phase structure of (\ref{genz}). 
In the next section we will discuss the phase structure of (\ref{standardmodel}).

\section{Critical behavior in matrix model}
 The eigenvalues of an unitary matrix $U$ are the complex numbers $e^{i\theta_i }$.\footnote{ Phase structure of a generic unitary matrix model has been discussed in \cite{Mandal:1989ry}.} In the large $N$ limit, we can consider an eigenvalue density $\rm \rho(\theta)$ defined on the unit circle by,
 
\bea
\rho(\theta) &=& \frac{1}{N}\sum_{i=1}^{N}\delta (\theta -\theta_i)= \sum_{n}\exp(i n\theta)\frac{1}{N} \tr \, U^n
\eea
The density function is non-negative and normalized,
\bea
\int \rm \rho(\theta) d\theta =1 \\
\rm \rho(\theta) \ge 0
\eea
It is well known that in the limit of $N\rightarrow \infty$, $\rho (\theta )$ can develop gaps, i.e. it can be non-zero only in bounded intervals. For example, in the case of a single gap when $\rho (\theta )$ is non-zero only in the interval $(-{\theta_0 \over 2},\frac{\theta_{0}}{2})$, it is given by the classical formula 

\bea
\rho(\theta)= f(\theta)\sqrt{\sin^2 {\theta_0 \over 2} -\sin^2{ \theta \over 2}}
\label{gapdistri}
\eea

A well known example of a $\rho(\theta)$ which does not have a gap is

\bea
\rho(\theta)= \frac{1}{2\pi}(1+a\cos(\theta)), \quad a<1
\label{ungapdistri}
\eea

At $a=1$, $\rho(\pi)=0$, and a gap will begin to open. For $a>1$ the functional form of $\rho(\theta)$ is as given by (\ref{gapdistri}).
\par

The matrix model under investigation has a complicated effective action. The saddle point distribution of the eigenvalues of the matrix $U$ may or may not have a gap, depending on the values of parameters $g_k$ in (\ref{standardmodel}). 
In the large $N$ expansion, the functional dependence of $F(g_k ,\bar g_k )$ on $g_k,\bar g_k$  depends on the phase, and we quote from the known results \cite{Goldschmidt:1979hq,Periwal:1990gf,Periwal:1990qb,CMD},

\bea
\label{Nexpansion}
N^2 F(g_k,\bar g_k) &=& N^2 \sum_k \frac{kg_k \bar g_k}{4} + e^{-2Nf(g_k,\bar g_k)}\sum_{n=1}^{n=\infty}\frac{1}{N^n}F_{n}^{(1)}, \quad {\rm ungapped} \\
\nn N^2 F(g_k,\bar g_k) &=& N^2 \sum_k \frac{kg_k \bar g_k}{4}+\sum_{n=0}^{n=\infty} N^{-\frac{2}{3}n}F_{n}^{(2)}, \quad g-g_c\sim o( N^{-\frac{2}{3}}) \\
\nn  N^2 F(g_k,\bar g_k) &=& N^2 G(g_k,\bar g_k)+\sum_{n=1}^{n=\infty}\frac{G^{(n)}}{N^2},\quad {\rm gapped}
\eea

In the above, we have assumed for simplicity that the eigenvalue distribution has only one gap. (In principle we can not exclude the possibility of a multi gap solution. But in this paper, since we are interested in the critical phenomena that results when the gap opens (or closes) we will concentrate on the single gap solution.) Near the boundary of phases, the functions $F_{n}(g)$ and $G_{n}(g)$ diverge. It is well known that in the leading order $N$, $F(g_k,\bar g_k)$ has a third order discontinuity at the phase boundary. This non-analytic behavior is responsible for the large $N$ GWW type transition. In the $o(N^{-\frac{2}{3}})$ scaling region near the phase boundary (the middle expansion in (\ref{Nexpansion})) this non-analytic behavior can be smoothened by the method of double scaling. This smoothening is important for our calculation of the double scaled partition function near the critical surface. \par
In (\ref{Nexpansion}) $f(g_k,\bar g_k),F_{n}^{(1)},F_{n}^{(2)}$ and $G^{n}(g_k,\bar g_k)$ are calculable functions using standard techniques of orthogonal polynomials. 
As an example, $G(g_k,\bar g_k)$ can be expressed as,
\be
G(g_k,\bar g_k)={1\over N}\log h_0 + \int_0^1 d\xi (1-\xi) \log\, f_0(\xi)
\ee
where  $f_0(\xi)$ and $h_{0}$ are determined in terms of $g_k,\bar g_k$ by a recursion relation of orthogonal polynomials for the unitary matrix model. 
It should be noted that in the ungapped phase all perturbative ($\frac{1}{N^2}$) corrections to the leading free energy vanishes. This follows from the fact that in the character expansion (strong coupling expansion) the ungapped free energy becomes an exact result. We also note that at $g_k= 0 = \bar g_k$, $f=0$ and the non-pertubative term is absent. 

\subsection{Gap opening critical operator at m=1 critical point}
We now derive the form of the critical operator that opens the gap and corresponds to the scaling region of width $o(N^{-\frac{2}{3}})$.      
 
\par
From (\ref{Nexpansion}) we can easily find the density of eigenvalues in the ungapped phase. 
\bea
\rho(\theta)&=&\frac{1}{2\pi}(1+\sum_{k}(k g_k\exp(ik\theta)+ k \bar g_k\exp(ik\theta)) \\
\nn \rm and \quad \rho_k&=&k g_k
\label{rhoungap}
\eea 
For a set of real $g_k$, the lagrangian (\ref{standardmodel}) is invariant under $U \rightarrow U^{\dagger}$. We will assume that the gap opens at $\theta=\pi$ according to $\rho(\pi - \theta)\sim (\pi - \theta)^2$, which characterizes the first critical point\footnote{In general the mth critical point is characterized by $\rho(\pi - \theta)\sim (\pi - \theta)^{2m}$.}. At the boundary of the gapped-ungapped phase (critical surface) we have $\rho(\pi)=0$. In terms of the critical fourier components $\rho^{c}_k$, it is the equation of a plane with normal vector $\tilde D_k = (-1)^k$
\bea
\sum_{k=-\infty}^{\infty} (-1)^k (\rho^{c}_k+\bar \rho^{c}_k)= -1 
\label{critsur1}
\eea
Now since $\rho^{c}_k = kg^{c}_k$ (up to non-perturbative corrections),
we get the equation of a plane 
\bea
\sum_{k=-\infty}^{\infty} (-1)^k k (g^{c}_k+\bar g^{c}_k)= -1 
\label{critsur}
\eea
where $g^{c}_k$ are the values of $g_k$ at the critical plane. Since the metric induced in the space of $g_k$ from the space of $\rho_k$ is $G_{k,k'} = k^2\delta_{k,k'}$, the vector that defines this plane is 
\be
C_k = \frac {(-1)^k}{k}
\ee
We mention that the exact values of $g^c_k$ where the thermal history of the small blackhole intersects the critical surface are not known to us as we do not know the coefficients of the effective lagrangian. However this information, which depends on the details of dynamics, does not influence the critical behavior. The information where the small blackhole crosses the critical surface is given by the saddle point equations (\ref{saddleqs1}), which are in turn determined by the $o(N^2)$ part of the partition function. 
\par
 Below we will show that the critical behavior is determined by the departure from the critical surface and not on where the thermal history intersects it, and conclude that the $o(1)$ part of the doubled scaled partition function is always determined in terms of the solution of the Painlev\'e II equation.  
\par
If we go slightly away from the critical surface by setting $g_k = g_k^c + \delta g_k$ and $\bar g_k = g_k^c + \delta \bar g_k$, then the gap opens provided $\rho (\pi) < 0$ \footnote{To calculate $\rho(\theta)$ we have used the ungapped solution in (\ref{Nexpansion})}. This condition is easily ensured by the choice $\delta g_k + \delta \bar g_k = tN^{-\frac{2}{3}}C_k$, $t < 0$, which is normal to the critical plane (\ref{critsur}). 

 The operator that corresponds to $\rho(\pi)=0$ at the first critical point is 
\bea
\hat O &=& 
\sum_{k=1}^{\infty} (g_k^c \tr \, U^{k}+\bar g_k \tr \, U^{\dagger k}) 
\label{scaleop}
\eea
 The gap at $\theta = \pi$ opens if we add a perturbation that leads to a small negative value for the ungapped solution of $\rho(\pi)$. Such a perturbation is necessarily in the direction of the vector $C_k$, because a perturbation that lies in the critical plane does not contribute to the opening of the gap. Hence we will set $(g_k-g_k^c) = N^{-\frac{2}{3}}\tilde t_k$. As we shall explain in Appendix A, $\tilde t_k = tC_k$, where $t = \tilde C \cdot \tilde {t}$ is an arbitrary parameter and $\tilde C$ is the unit vector corresponding to $C$. Therefore the $\textit{relevant}$ gap opening perturbation to be added to the action is 
\bea
\hat O_t &=& 
N^{-\frac{2}{3}}t\sum_{k=1}^{\infty} C_k(\tr \, U^{k}+\tr \, U^{\dagger k}) 
\label{scaleop_t}
\eea
The factor $N^{-\frac{2}{3}}$ is indicative that the perturbation is relevant and has exponent ${-\frac{2}{3}}$. $N$ acts like an infrared cutoff 
.  
\par
In the double scaling limit, near the critical surface, $F_{0}^{(2)}$ in (\ref{Nexpansion}) is a function of the parameter $t$ (see Appendix A). It is known that $F_{0}^{(2)}(t)$ (from now on we will call it $F(t)$) 
satisfies the following differential equation,
\bea 
\ddel{F}{t} =-f^2(t) 
\eea
where $f(t)$  satisfies the Painleve II equation,
\bea
\frac{1}{2}\ddel{f}{t}=tf+f^3
\eea
The exact analytic form of $F(t)$ is not known, but $F(t)$ is a smooth function in the domain $(-\infty,\infty)$. 
Smoothness of $F(t)$ guarantees the smoothening of large N transition in the double scaling limit.
\par
In the gapped phase of the matrix model, $F(g_k,\bar g_k)$ has a standard expansion in integer powers of  $\frac{1}{N^2}$, which becomes divergent as one approaches the critical surface. In the double scaling region (\ref{Nexpansion}) $(g-g_c) \sim \CO(N^ {- {2\over 3}})$, and the the perturbation series (\ref{Nexpansion}) is organized in an expansion in powers of $N^{-\frac{2}{3}}$. The reason for the origin of such an expansion is not clear from the viewpoint of the bulk string theory. However, it is indeed possible to organize the perturbation series, in the scaling region, in terms of integral powers of a renormalized coupling constant. We will come back to this point later. In the ungapped phase the occurrence of $o(e^{-N})$ terms is also interesting. Here too we lack a clear bulk understanding of the non-perturbative terms which naturally remind us of the D-branes.    

\section{Saddle point equations at large N}
In this section we will use the results of the previous section to write down the large $N$ saddle point equations for the multi-trace matrix model (\ref{fullint}). We treat $\mu_j$ and $\bar \mu_j$ as independent complex variables. This is natural as the saddle point of the theory may occur at complex values of the variable, though at the end we will find that for real $\alpha_{\vec k , \vec k'}$ in (\ref{genaction}) we have saddle points in imaginary $\mu_i$ and real $g_i$. From (\ref{genaction}) we deduce the saddle point at large $N$ by including the leading $o(N^2)$ contribution of $F(g_k,\bar g_k)$ to the free energy. The equations for saddle points are given by
\bea
\label{saddleqs1}
{\partial S_{\rm eff} \over \partial g_j}&=& i \bar \mu_j +  {1\over 2j} \bar g_j=0, \\
\nn {\partial S_{\rm eff} \over \partial \bar g_j} &=& i \mu_j  + {1\over 2j}  g_j=0 , \\
\nn {\partial S_{\rm eff} \over \partial \mu_j}&=&-a_j \bar \mu_j  + i \bar g_j + \sum_{\vec k, \vec k'} \alpha_{\vec k , \vec k'} (-{\rm i})^{|\vec k|+ |\vec k'|} {k_j' \over \mu_j} \Upsilon_{\vec k} (\bar \mu) \Upsilon_{\vec k'} (\mu)=0 \\
\nn {\partial S_{\rm eff} \over \partial \bar\mu_j}&=& -a_j \mu_j  + ig_j +\sum_{\vec k, \vec k'} \alpha_{\vec k , \vec k'} (-{\rm i})^{|\vec k|+ |\vec k'|} {k_j \over \bar \mu_j} \Upsilon_{\vec k}  (\bar \mu) \Upsilon_{\vec k'} (\mu)=0 
\eea
These equations correspond to the ungapped phase. Equations similar to equation (\ref{saddleqs1}) can also be written using $F(g_k,\bar g_k)$ in the gapped phase. 

\par
By the AdS/CFT correspondence the solutions to (\ref{saddleqs1}) are dual to supergravity/string theory solutions, like $AdS_5 \times S^5$ and various $AdS_5 \times S^5$ blackholes. The number and types of saddle points and their  thermal histories depends on the dynamics of the gauge theory (i.e. on the numerical values of the parameter $a_j$ and $ \alpha_{\vec k, \vec k'}$, which in turn are complicated functions of $\lambda$ and $\beta$). These issues have been discussed in the frame work of simpler models in \cite{Alvarez-Gaume:2005fv}, where the first order confinement/deconfinement transition and its relation with the Hawking-Page type transition in the bulk has also been discussed. Here we will not address these issues, but focus on the phenomenon when an {\it unstable saddle point} of (\ref{saddleqs1}) crosses the critical surface (\ref{critsur}). 
\par
By solving the eqn.(\ref{saddleqs1}) we can write $g_j$ in terms of $\mu_j$ and the coefficients $a_j(\beta),\alpha_{\vec k, \vec k'}(\beta)$. Using the critical values of $g_j$ (\ref{critsur}), we get the relation between $a_j(\beta),\alpha_{\vec k, \vec k'}(\beta)$ at the critical surface,
\bea
\label{criticalplane}
 g^c_j (j a_j -1) + {\hat g^c_j \over j} + \sum_{\vec k, \vec k'}  2^{2 -|\vec k| -|\vec k '|} (-1)^{|\vec k| +|\vec k '|}\alpha_{\vec k , \vec k'} {k_j \over g^c_j} \Upsilon_{\vec k +\vec k'}(g^c_j)=0, \quad j=1, \cdots, p.
\eea
Whether the above relation is achieved for some values of the co-efficients $a_j(\beta),\alpha_{\vec k, \vec k'}(\beta)$ is a difficult question which again needs a detailed understanding of the gauge theory dynamics. The coefficients $a_j(\beta),\alpha_{\vec k, \vec k'}(\beta)$ have been perturbatively calculated in \cite{Aharony:2005bq} and it can be shown that at some specific $\beta < \beta_{\rm HG}$ \footnote{$\beta_{\rm HG}^{-1}$ is the temperature of Hagedorn transition.} the condition (\ref{criticalplane}) is satisfied. 
\par
We would like to mention that there is no fine tuning associated with the relation (\ref{critsur}) or (\ref{criticalplane}) being satisfied. This is because we have one tunable parameter, the temperature, and one relation (\ref{critsur}) to satisfy. Hence one may hope that in the most general situation the relation (\ref{critsur}) will be satisfied. In the next section we will discuss the doubled scaled partition function near the critical point.   
\par
In a later section we will use the AdS/CFT correspondence to argue that in the strongly coupled gauge theory, a 10 dimensional ``small blackhole'' saddle point reaches the critical surface (\ref{criticalplane}). The interpretation of this phenomenon in the bulk string theory, as a blackhole to excited string transition will also be discussed. 

\section{Double scaled partition function at crossover}\label{finalsection}

We will assume that the matrix model (\ref{standardmodel}) has a saddle point which makes a gapped to ungapped transition as we change the parameters of the theory($\alpha_{\vec k, \vec k'}^c$,$a_j$) by tuning the temperature $\beta^{-1}$. We will also assume that, this saddle point has one unstable direction which corresponds to opening the gap as we lower the temperature. These assumptions are motivated by the fact that the small (euclidean) Schwarzchild blackhole crosses the critical surface and merges with the $AdS_5 \times S^5$ and that it is an unstable saddle point of the bulk theory. To calculate the doubled scaled partition function near this transition point, we basically follow the method used in \cite{Alvarez-Gaume:2005fv}. We expand the effective action (\ref{standardmodel}) around the $1{\it st}$ critical point, and we simultaneously expand the original couplings $a_j$, $ g_j$, $\bar g_j$ and $\alpha_{\vec k, \vec k'}$ around their critical values  $a_j^c$, $\beta_j^c$, $g_j^c=0$, and $\alpha_{\vec k, \vec k'}^c$. For clarity we define  
\bea
P(\mu, \bar \mu, \alpha)=\sum_{\vec k, \vec k'} \alpha_{\vec k , \vec k'} (-{\rm i})^{|\vec k| + |\vec k'|} \Upsilon_{\vec k} (\bar \mu) \Upsilon_{\vec k'}(\mu)
\eea

 We also introduce the column vectors,

\be
\label{varias}
 \mu =  \left( \begin{matrix}  \mu_j \\
                                  \bar\mu_j \end{matrix} \right), \quad
 A= \left( \begin{matrix} a_j \\
                                  \alpha_{\vec k, \vec k'}   \end{matrix} \right), 
                             \quad
                             g= \left( \begin{matrix} g_j \\
                                  \bar g_j \end{matrix} \right)                                   
\ee
and expand the above mentioned vector variables
\bea
 g- g^c &=& N^{-\frac{2}{3}}\tilde{t} \\
\nn \mu- \mu^c &=& N^{-{4 \over 3}} n \\
\nn  A - A^{c} &=& \tilde{g}  N^{-\frac{2}{3}} \alpha
\label{scaling}
\eea
where $\tilde{g}=N^{\frac{2}{3}}(\beta-\beta_c)$ and $\alpha={\partial A \over \partial \beta}|_{\beta=\beta_c}$.  The expansion of the co-efficients $a_j$ and $\alpha_{\vec k, \vec k'}^c$
 are proportional to the deviation of the tuning parameter $\beta$ from its critical value, i.e. $\tilde{g}=N^{\frac{2}{3}}(\beta_c-\beta)$.

The expanded action takes the following form,
\be
N^2 S_{eff}=-{1\over 2} N^{-\frac{2}{3}} n^t\, {\cal L} \, n +  n^t({\cal J} t - \tilde{g} {\cal H} \alpha) + F(C \cdot \tilde{t})+O(N^{-\frac{4}{3}}) 
\ee

In the above we have, following the discussion in Appendix A, used the  fact that the $o(1)$ function $F$ depends on the scaled variable through the combination $t=C \cdot \tilde{t}$. Recall that $C$ is the constant vector normal to the critical plane and the matrices ${\cal L}$, ${\cal J}$, ${\cal H}$ are given by

\be
\begin{aligned}
{\cal L}= &\left( \begin{matrix}  -{\partial ^2 P \over \partial \mu_j \partial \mu_k} & a_j^{(c)}\delta_{jk} -  {\partial ^2 P \over \partial \mu_j \partial \bar \mu_k}\\
                               a_j^{(c)}\delta_{jk} -  {\partial ^2 P \over \partial \mu_j \partial \bar \mu_k}   &   -{\partial ^2 P \over \partial \bar \mu_j \partial \bar \mu_k} \end{matrix} \right), \\ 
{\cal H}=  &\left( \begin{matrix}  - \bar \mu_j \delta_{jk} & {\partial ^2 P \over \partial \mu_j \partial \alpha_{\vec k, \vec k'}}\\
                               -\mu_j \delta_{jk}   & {\partial ^2 P \over \partial \bar \mu_j \partial \alpha_{\vec k, \vec k'}} \end{matrix} \right), \\ 
 {\cal J}=& {1\over 2} \left( \begin{matrix}  {\rm i} {\cal F} & {\cal F} \\
                               {\rm i} {\cal F}   & -{\cal F} \end{matrix} \right),
                               \end{aligned}
                               \ee
In the above we have introduced the diagonal matrix 
\be
{\cal F}_{jk}={1\over j} \delta_{jk}, \quad j,k=1, \cdots, p.
\ee
All quantities appearing in the matrices are calculated at the first critical point. Here $o(N^2)$ part of the action does not depend on $n,\tilde t$ and hence we do not show this part of the action explicitly.

We now do the Gaussian integration over $n_k$ in the functional integral
\be
\label{expg}
Z \sim \int d\tilde t ({\rm det} ({N^{-\frac{2}{3}}\cal L}))^{-{1\over 2}} \exp \biggl\{ {1\over 2} N^{\frac{2}{3}} (\tilde t- \tilde{g} {\cal C} \alpha)^t {\cal M} (\tilde t - \tilde{g} {\cal C} \alpha) + 
 F(C \cdot \tilde t) + O(N^{-\frac{2}{3}})\biggr\},
\ee
The matrices appearing here can be easily obtained,

\be
{\cal D}={1\over 2} \left( \begin{matrix}  {\cal F} & 0 \\
                                 0 & {\cal F}  \end{matrix} \right), \quad
{\cal M}={\cal J}^t {\cal L}^{-1} {\cal J} +{\cal D} ,\quad
{\cal C}=-{\cal M}^{-1} {\cal J}^t {\cal L}^{-1} {\cal H}.
\ee
Notice that the Hessian associated with $S_{\rm eff}$ is given by
\be
H=  \left( \begin{matrix} - {\cal L} & {\cal J} \\
                                {\cal J}  & {\cal D}  \end{matrix} \right).
\ee
\par
In order to discuss the further evaluation of the integral (\ref{expg}),
we must take into account the fact that we are evaluating the integral near an unstable saddle point. That the saddle point has precisely one unstable direction is motivated by the fact that in the bulk theory the euclidean 10-dimensional blackhole has one negative eigenvalue.
This statement strictly speaking should apply to the saddle point in the gapped phase. However since the GWW phase transition is third order  
an unstable saddle point in the gapped phase should continue to be unstable at the crossover. 

In order to render the gaussian integral (\ref{expg}) along the unstable direction well defined, we should make an analytic continuation. Once this is done we can easily see that as $N\rightarrow \infty$ the integral in (\ref{expg}) is localized at 
\be
\tilde{t}=\tilde{g} {\cal C} \alpha
\ee
This follows from a matrix generalization of the gaussian representation of the delta function.  

Putting the above expression in ({\ref{expg}) we get the final result,
\bea
Z \sim  i({\rm det}(H))^{-{1\over 2}}\exp F(\tilde{g} C \cdot {\cal C}\alpha),
\label{finalresult}
\eea
where  $C \cdot {\cal C}{\alpha}$ is a constant independent of $\tilde{g}$. We have assumed that the Hessian $H$ does not have a zero mode, but the one negative eigenvalue accounts for the $i$ in front of (\ref{finalresult}). 
  \par
 The $o(1)$ part of the partition function, (\ref{finalresult}) is universal in the sense that the appearance of the function $F(\tilde{g} \times constant)$, does not depend on the exact values of the parameters of the theory. In the double scaling limit the partition function becomes a function of a single scaling variable $\tilde{g}$. Exact values of the couplings and the $o(N^2)$ part of the partition function determine where the thermal history crosses the critical surface (\ref{critsur}). However the form of the function $F$ and the double scaling limit of (\ref{scaling}) are independent of the exact values of $g^{c}_k$. They only depend on the fact that one is moving away perpendicular to the critical surface. This is the reason why in \cite{Alvarez-Gaume:2005fv} we obtained exactly the same equation when $g^{c}_1\neq 0$ but all other $g^{c}_k=0$. 
\subsection{Condensation of winding modes at the crossover}

We will now discuss the condensation of the winding Polyakov lines in the crossover region. Specifically we will discuss the expectation value of the critical operator (\ref{scaleop}). In the leading order in large N we have already seen in (\ref{rhoungap}), that $\rho_k^c = kg_k^c$. 
In order to calculate subleading corrections it can be easily seen that all the $\rho_k$'s condenses in the scaling region, 
\bea
\vev{N^{\frac{2}{3}}(\rho_k-\rho^{ug}_k)} = C_k \frac{dF}{dt}
\label{condensation}
\eea
where $\rho^{ug}_k=kg_k$.  
 This smoothness of the expectation value of the $\rho_k$'s follows from the smooth nature of $F(t)$. The exact form of $F(t)$ is not known but it is known that it is a smooth function with the following asymptotic expansion. 
\bea
 F(t) &=&  {t^3 \ov 6}
 - {1 \ov 8} \log (-t) - {3 \ov 128 t^3}
 + {63 \ov 1024 t^6} + \cdots  ,\quad -t \gg 1 \\
\nn F(t) &=& {1 \ov 2 \pi} e^{-  {4 \sqrt{2} \ov 3} t^{3 \ov 2}}
 ( - {1 \ov 8 \sqrt{2} t^{3 \ov 2} } + {35 \ov 384 t^3} -
 {3745 \ov 18432 \sqrt{2} t^{9 \ov 2}} + \cdots ), \quad  t \gg 1 
 \eea

The derivative of $F(t)$ diverges as $t \rightarrow -\infty$ and goes to zero as $t \rightarrow \infty$. This behavior tallies with the condensation of winding mode in one phase (the gapped phase) and the non-condensation of winding modes in the ungapped phase. The condensation of the winding modes also indicates that the $U(1)$ symmetry (which is the $Z_N$ symmetry of the $SU(N)$ gauge theory in the large N limit) is broken at the crossover, but restored in the limit $t \rightarrow \infty$.

\section{Higher critical points and the introduction of chemical potentials}\label{multcrit}
Besides the first critical point, single trace unitary matrix models
can have higher critical points. The m${\it th}$ critical point is
characterized by,
\be
\rho_m(\theta) \sim (\theta-\pi)^{2m}, \quad \theta \rightarrow \pi,
\ee
and hence it is specified by the following relations,
\bea
\rho^{(2n)}(\pi)=0, \quad 0 \le n < m
\eea
Writing the above in terms of $g_k$'s we get,
\bea
\sum_{k=-\infty}^{\infty}(-1)^{k}k^{2n-1}(g^{c}_k+\bar g^{c}_k)=0,
\quad 0 \le n < m
\label{newmulticrit}
\eea

A particular choice for the density of eigenvalues with this
behavior is
\be
\rho_m(\theta)=c_m \biggl( 2\,\cos \, {\theta \over 2}\biggr)^{2m},
\ee
where
\be
c_m = {2^{2m}\over 2\pi} {(m!)^2\over (2m)!}.
\ee
By expanding in Fourier modes, one finds
\be
\rho_m(\theta)={1\over 2\pi}\biggl( 1 +2\sum_{k=1}^{m}  { (m!)^2
\over (m-k)!(m+k)!}\cos\, k\theta \biggr)
\label{newps}
\ee
   Using the relation between the density of eigenvalues in the
ungapped phase and the matrix model potential one recovers the
critical potential of Periwal and Shevitz.

As the plane (\ref{newmulticrit}) is determined by more than two
equations, a generic curve in the space of couplings $g_k$'s will not
necessarily intersect the plane. Hence by tuning one parameter, the
history of a saddle point may not reach the higher critical points.
But one may consider a situation where along with temperature, some
additional chemical potentials are also turned on \cite{jaffe}. Using  these
chemical potentials (like say the R-charge) we may be able to reach
higher multicritical points.
\par
In appendix (\ref{app:multicritical}), we have considered a more general 
effective action which includes general source terms in addition to
(\ref{genaction}),
\bea
\widetilde S(U, U^{\dagger})=S (U, U^{\dagger})
+ N \sum_{k\ge 1} ( b_k {\rm tr}\, U^k + {\bar b}_k  {\rm tr}\, U^
{\dagger k}).
\label{newmulact}
\eea
Using the above action, we have calculated the doubled scale
partition function near higher critical points. Similar to our result in (\ref
{finalresult}), the $o(1)$ part of the doubled scaled partition
function becomes a universal function determined by the mKdV
hierarchy. It should be mentioned that the calculation is performed near the m--th multicritical point characterized by,
\be
g_n=0 \quad, n>m
\label{choice}
\ee
According to the comments at the end of section(\ref{finalsection})
the final form of the doubled scaled partition function(\ref
{newfinalresult}) and the double scaling limit (\ref{newrelcrit}) is
universal and independent of the particular choice of (\ref{choice}).

\section{Applications to the small 10-dimensional blackhole}
We now apply what we have learned about the matrix model (gauge theory) GWW transition and its smoothening in the critical region to the blackhole-string transition in the bulk theory. The first step is to identify the matrix model phase in which the blackhole or for that matter the supergravity saddle points occur. We will argue that they belong to the gapped phase of the matrix model. This inference is related to the way perturbation theory in $\frac{1}{N}$ is organized in the gapped, and ungapped phase as discussed in (\ref{Nexpansion}). Note that it is only in the gapped phase, that the $\frac{1}{N}$ expansion is organized in powers of $\frac{1}{N^2}$, exactly in the way perturbation theory is organized around classical supergravity solutions in closed string theory. Hence at the strong gauge theory coupling($\lambda \gg 1$), it is natural to identify the small 10 dimensional blackhole with a saddle point of the equations of motion like (\ref {saddleqs1}) but obtained by using $F(g_k,\bar g_k)$  corresponding to the gapped phase. \footnote{A saddle point of the weakly coupled gauge theory may also exist in the gapped phase. With a change in the temperature the saddle point can transit through the critical surface. Using the results of \cite{Alvarez-Gaume:2005fv}, it is easy to see that this is precisely what happens for the perturbative gauge theory discussed in \cite{Aharony:2005bq}. We note that in the corresponding bulk picture since $l_s >> R_{AdS}$, the supergravity approximation is not valid. It would be interesting to understand the bulk interpretation in this case.} One can associate a temperature with this saddle point which would satisfy $l_s^{-1} \gg T \gg R^{-1}$. 

As the temperature increases towards  $l_s^{-1}$, one traces out a curve (thermal history) in the space of the parameters $a_i, \alpha _{k, k'}$ of the effective theory. 
One can also say that a thermal history is traced in the space of $\rho _i =\langle \frac{1}{N}\tr \, U^i\rangle$, which depends 
on the parameters of the effective theory. We will now make the reasonable assumption that the thermal history, at a temperature $T_c \sim l_s^{-1} $, 
intersects the critical surface (\ref {critsur1}) (equivalently the plane (\ref {critsur}) and then as the temperature increases further it reaches the point 
$\rho _i =\langle \frac{1}{N}\tr \, U^i \rangle = 0$, which corresponds to $AdS_5\times S^5$. Once the thermal history crosses the critical surface, the gauge theory 
saddle points are controlled by the free energy of the ungapped phase in (\ref{Nexpansion}). The saddle points of eqns. (\ref{saddleqs1}) which were obtained 
using this free energy do not correspond to supergravity backgrounds, because the temperature, on crossing the critical surface is very high $T \gtrsim l_s^{-1}$. Besides 
this the free energy in the gapped phase has unconventional exponential factors (except at $g_k=0$ which corresponds to $AdS_5\times S^5$). It is likely that these 
saddle points define in the correspondence, exact conformal field theories/non-critical string theories in the bulk. Neglecting the exponential corrections $exp(-N)$, it seems 
reasonable, by inspecting the saddle point equations, that in this phase the spectrum would be qualitatively similar to that around $\rho _i = 0$. Since this corresponds to 
$AdS_5\times S^5$, we expect the fluctuations to resemble a string spectrum.

As we saw in the previous section, our techniques are good enough only to compute the $o(1)$ part of the partition function in the vicinity of the critical surface which depends on the renormalized coupling. The exact solution of the free energy (in the single trace model) in the transition region in (\ref{Nexpansion}) enabled us to define a double scaling limit  
in which the non-analyticity of the partition function could be smoothened out, by a redefinition of the string coupling constant according to 
$\tilde{g}=N^{\frac{2}{3}}(\beta_c-\beta)$. This smooth crossover corresponds to the blackhole crossing over to a state of strings corresponding to the ungapped phase. 

We have also computed the vev of the scaling operator and hence at the crossover the winding modes $\rho _i = \langle \frac{1}{N}\tr \, U^i\rangle$ condense (\ref{condensation}). They also 
have a smooth parametric dependence across the transition. This phenomenon in the bulk theory may have the interpretation of smooth topology change of a 
blackhole spacetime to a spacetime without any blackhole and only with a gas of excited string states. However in the crossover region a geometric spacetime 
interpretation is unlikely. We may be dealing with the exact description of a non-critical string in 5-dims. in which only the zero mode along the $S^3$ directions 
is taken into account. This interpretation is inspired by the fact that the free energy $F(t)$ also describes the non-critical type 0B theory as was already discussed in \cite{kms,Alvarez-Gaume:2005fv}.

\section{Implications for the Lorentzian blackhole, the information puzzle, and related matters}
All our discussion has been in the context of the euclidean time, both in the bulk and the boundary theory. Since the boundary theory is governed by a well defined positive Hamiltonian the analytic continuation from euclidean to lorentzian signature is well understood and simple. Hence the partition function gives a way of computing the density of states at a particular energy using the formula,
\bea
Z(\beta)=\int_{0}^{\infty}dE \rho (E)e^{-\beta E}
\label{densityE}
\eea
where $\rho(E) = \tr \, \delta (H - E)$ is the density of states at energy $E$. Since the partition function, in an appropriate scaling limit, is a smooth function of the renormalized coupling constant $\tilde g$, at the crossover between the gapped and ungapped phase, (\ref{densityE}) implies that $\rho(E)$ inherits the same property. Since $\rho(E)$ is as well a quantity that has meaning when the signature of time is Lorentzian, it would imply that the blackhole-string crossover in the Lorentzian signature is also smooth. This is an interesting conclusion especially because we do not know the AdS/CFT correspondence for the small Lorentzian blackhole. The Lorentzian section of the blackhole has a horizon and singularity. Since the gauge theory should also describe this configuration, a smooth density of states in the cross over would imply that the blackhole singularity was resolved in the gauge theory.
\par
 We believe in this conclusion but an understanding of this can only be possible if we have an explicit model in the gauge theory of the small Lorentzian blackhole. Work in this direction is in progress drawing lessons from \cite{Witten:1998zw,Aharony:2005bm,Maldacena:2001kr,Kraus:2002iv,Fidkowski:2003nf,Festuccia:2005pi}.

\par 
This program was originally motivated by an attempt to understand and resolve the information puzzle in blackhole physics. In the AdS/CFT correspondence we know that the $SU(N)$ gauge theory is defined by a hermitian hamiltonian defined on $S^3\times R$. The $N\rightarrow \infty$ limit and the $\lambda \rightarrow \infty$ limits make contact with semi-classical gravity limit of the type IIB string theory in the bulk. In this limit, one can represent the quantum gravity theory path integral as an integral which splits into a sum over distinct topologies. In particular in the euclidean framework the path integral splits as a sum of contributions from histories with and without a blackhole. However this representation arises by a naive consideration of the large N limit. We know that as long as N is finite the notion of summing over distinct topologies does not exist. A careful understanding of the double scaling limit has indeed made it possible to treat finite N effects in a saddle point expansion around large N and smoothened the GWW transition.
Since we have identified this gauge theory phenomenon with a smooth blackhole-string crossover, we conclude that topology change is indeed possible in the bulk string theory.

In light of our results we are not convinced about Hawking's proposed solution to the information puzzle \cite{Hawking:2004wm} which uses the notion of representing the quantum gravity path integral as a sum over all topologies. At large but finite N (or equivalently at small but finite string coupling) this notion is not neccessarily valid.



\section{Appendix A: Discrete recursion relations, $m=1$ critical point and Painleve II}{\label{app:Painleve}}

In this appendix we discuss the appearance of the m=1 critical point
in the discrete recursion relations in the presence of general
couplings $g_k$, where k is a positive integer. The main point can be
explicitly illustrated in the case of two couplings $g_1$ and $g_2$,
and the generalization to more general potentials is straightforward.
We briefly review how we find scaling regions in matrix models and
how double scaling limits are implemented.
We follow closely the work of Periwal-Shevitz  \cite{Periwal:1990gf}.
The  action we consider is:
\be
g_1 \Bigl( \tr\, U + \tr\, U^{\dagger}\Bigr)+ g_2 \Bigl( \tr\, U^2 +
\tr\, U^{\dagger 2}\Bigr) =
\mu_1 V_1 + \mu_2 V_2,
\ee
where $V_{1,2}$ are the  first critical potentials found in \cite
{Periwal:1990gf}:
\be
\ba
V_1 ={1\over 2} \Bigl( \tr\, U + \tr\, U^{\dagger}\Bigr), \\
V_2={4\over 3} \Bigl( \tr\, U + \tr\, U^{\dagger}\Bigr) +{1\over 12}
\Bigl( \tr\, U^2 + \tr\, U^{\dagger 2}\Bigr).
\ea
\ee
and
\be
\mu_1=2g_1- 16 g_2, \qquad \mu_2=12 g_2.
\ee

For those interested in the details we  have  modified the  critical
potentials by making the transformation
$g_k\rightarrow (-1)^k g_k, \,\, U\rightarrow -U$.  This is  a
symmetry of the action that guarantees that
the gap opens at $\theta=\pi$.  In the original paper \cite{Periwal:1990gf} the gap opens at $\theta=0$.
Obviously the  gap can open anywhere  on the circle, but  we simply
have to be consistent once a convention
is chosen.
The Periwal-Shevitz's \cite{Periwal:1990gf} equation with two
couplings $g_1$ and $g_2$, in our convention takes the following form,
\bea
-R_n \frac{n+1}{N}=(1-R_n^2)[-(R_{n+1}+R_{n-1})g_1-2g_2(R_{n-1}R_{n-1}
^2+\\
\nn R_{n-1}^2R_{n}+2R_{n-1}R_{n}R_{n+1}+R_{n}R_{n+1}^2-R_{n+2}-R_{n-2}
+R_{n+1}^2R_{n+2})]
\label{recursion}
\eea

We will show that this equation besides the m=2 fixed point also has
the m=1 fixed point. The latter is well known to be described by
Painleve II equation with just one coupling. (The derivation of
Painleve II from the one coupling case has been discussed in the
original paper \cite{Periwal:1990gf}).

As usual  to find scaling regions we first solve the planar theory.
However we have
to solve it for any $n$, in other words, in the planar case $R_n$
becomes a function
$R(\xi)$, where $\xi=n/N$ which completely determines the planar
limit of the  theory.
The equation that determines
$R(\xi)$ is  obtaining by ignoring in \ref{recursion} the above the
shifts in the $R's$.  This yields
the planar  string equation:
\be
R\xi = (1-R^2)( 2(g_1-2 g_2) R + 12 g_2 R^3)
\label{saddle}
\ee
If we take the scaling region to be close to the endpoint of the $\xi
$ interval, i.e. $1$,
we introduce the scaling variable:
\be
\xi= 1-a^2 t
\ee
as is standard in matrix models, and $a$ is a small ``lattice"
parameter that is necessary
to study the  scaling region.  Since in these theories the  critical
value of $R=0$, we have
to write the function $R$ in terms of some scaling funcion with
appropriate exponents:
\be
R= a^{\gamma} f(t)
\ee
Since we want to consider only the first critical point $m=1$, this
implies that $\gamma=1$
and the scaling behavior of $R$ is
\be
R=a f(t)
\ee
Substituting in the planar string equation we obtain:
\be
a f (1-a^2 t)= (1-a^2 f^2) ( 2 (g_1-2g_2) a f +12 g_2 a^3 f^3)
\ee
The terms of order $a$ determine the criticality condition, which as
expected is the gap opening
condition
\be
g_1-2 g_2=\frac{1}{2}
\label{mone}
\ee
The terms of order $a^3$ now provide the planar string  equation that
determines the functional
form of $f$ as a function of $t$ to leading order in $1/N$:
\be
-a^3 t f(t)=-a^3 f^3 (2 (g_1-2g_2)-12 g_2)
\label{planarstring}
\ee
all other terms are irrelevant to this order, and what this  equation
does is to determine
$f(t)$, and also it provides the  first term in the expansion of the
P-II equation in powers
of fractional powers of $t$.
The condition \ref{mone} determines the first critical point of the
theory, $m=1$, which implies
that near $\xi=1$ equation \ref{saddle} has a second order zero in
R.  If we require that the
zero is of order $4$ (after dividing by a common $R$ on both sides)
we obtain the conditions
for the $m=2$ critical point  governed by the scaling action $V_2$
above.  Since  in our  problem
we have a single control paremeter, i.e. the temperature, we focus on
the  $m=1$ condition
\ref{mone} and study next the  double scaling limit.  To make contact
with the arguments of section
VI we will study this limit for generic coupling $g_1,\, g_2$, this
way we include also the perturbations
of a given model on the ``critical surface" \ref{mone} by the  gap
opening operator (VI.44).

So far the parameter $a$ is just a small number, and for the time
being it has no dependence
on $N$.  To get the  $N$-dependence we do the double scaling limit,
by expanding the full string equation,
and see what is the relation beteween $N$ and $a$ that leads to a
differential equation containing
the string coupling constant, i.e. containing higher genus terms in
the expansion and thus
generating  a string perturbation theory.  Let us  do it  in general,
but of course we have
to keep track of the  fact that we have already determined the
scaling behavior of both
$\xi$ and $R(\xi)$, and we have to include it in \ref{recursion}:
\bea
a f (\xi)(1-a^2 t)\, &=&\left( 1 - a^2\,{f(\xi)}^2 \right) \,
    \left( 2\,a\,{g_1}\,f(\xi) -
      4\,a\,{g_2}\,f(\xi) +
      12\,a^3\,{g_2}\,{f(\xi)}^3 \right) \\
\nn &+ & \left( 1 - a^2\,{f(\xi)}^2 \right) \,
    \Bigl( 20\,a^3\,{g_2}\,f(\xi)\,
       {f'(\xi)}^2 \\
     \nn  &+&\,  a\,({g_1}-8g_2)\,f''(\xi) +
      20\,a^3\,{g_2}\,{f(\xi)}^2\,f''(\xi)
     \Bigr) \,{1\over N^2} + \ldots
\eea
Now we are ready to get the relation between $N$,  and $a$.  In going
from derivatives with respect to $\xi$
to derivatives with respect to $t$, we obtain, including the factor
of $1/N$ a term of  the form:
\be
{1\over N a^2}{d\over d t}
\ee
for each derivative.  Since the  first nontrivial terms with
derivatives contains two of them, this
means:
\be
   {1\over ( N a^2)^2}{d^2\over d t^2}
   \ee
The final result up to two derivatives (it is easy to show that
higher ones are irrelevant) is:
\bea
\label{finalequation}
   -a^3 t f(t) &=& -(1-12 g_2) a^3 f(t)^3   \\
    \nn  &+&\, (g_1-8 g_2) a  {1\over ( N a^2)^2}{d^2 f \over d t^2}\\
   \nn  &+&\, 20\, g_2 a^3 {1\over ( N a^2)^2}( f \dot{f}^2+f^2 \ddot
{f})
\eea
where the dots are derivatives with respect to $t$.  To get the
double scaling limit,
notice that we want that up to a numerical constant
\be
a( {1\over N a^2})^2= g_{st}^2 a^3
\ee
Hence, up to $g_{{\rm st}}$ we obtain:
\be
a\sim N^{-1/3}
\ee
Note that the terms in the third line of \ref{finalequation}
will vanish like $a^2$ after we divide out by $a^3$
unless we force a strange scaling of $g_2$, but this is something we
cannot do
in the above procedure.  The equation that survives is of course
Painleve-II after
some simple numerical rescalings
The computation has been carried out  only for the two coupling case,
but  it is easy to
generalize to a more general action.  We have also included the case
where we
have a shift of the  couplings of the model with respect to the
critical surface.  Of course
the answer is the same, and the reason is that any of the terms $\tr
(U^k+U^{-k})$
that appear in the gap opening operator have a component along  the
first scaling
operator.  For the  two coupling theory this is the  origin of  the
term $-12 g_2$ in the $f^3$ piece
and the term $-8g_2 a$ in the term $\ddot{f}$.  We get Painleve-II
unless we do
some unnatural fine tuning in the coupling $g_2$, a freedom we do
not  have at our
disposal given that we have just one  control parameter.
Obviously, even if  we consider more
general  potential, the same will happen with the gap opening operator.
The operator identified with gap opening in the text should be more
precisely
be called the ``bare" gap opening operator.  After
renormalization around any critical point, and in particular near the
$m=1$ it will be be
dominated by the  first scaling operator.  We know also from \cite
{Periwal:1990gf}
that the integrable  hierarchy behind the unitary matrix model is
the  modified
KdV (mKdV), and their flows can be identified with the expectation
values  of the
scaling  operators of the theory (including of course the irrelevant
ones at the
$m=1$ critical surface.

One may wonder what happens with the expectation values of the
$\rho_n$ at the cross over region.  This is  however no problem,
since we can
renormalize these operators with more freedom than we have above, in
fact, the way
to argue that generically, at the initial conditions of the mKdV
hierarchy that  starts
with Painleve-II; the continuum limit of the $\rho_n$ get an
expectation value is to use the
renormalized  Wilson loop operator of the  matrix model, as it is
done in \cite{CMD}.
The expansion of the Wilson loop $\langle w(t)\rangle$ has as
coefficients, for each power of $t^{n+1}$ precisely the expectation
value of the
corresponding $\sigma_n$ which are the continuum limits of the $\rho_n
$, and
what follows from the double scaling limit of the  loop equations is
that to leading order those expectation values are not
zero and are given by a power of $f$ to leading planar order with
corrections.
This power of course is not zero, and hence it says  that the
corresponding  derivative
of the free energy with respect to the scaling parameter $t_n$ that
produces the
expectation value of $\rho_n$ is not zero even when we set $t_n=0$
after taking
the derivative.


\section{Appendix B: Partition function near multicritical points}{\label{app:multicritical}}
Here we will calculate the double scaled partition function near higher multicritical points. We start with eqn (\ref{newmulact}) and denote
\be
\label{newgenz}
Z=\int [d\, U] e^{\widetilde S(U,U^{\dagger})},
\ee
where $\widetilde S(U,U^{\dagger})$ has the form (\ref{newmulact}). We will assume in the following that $a_i>0$. We closely follow the discussion of section \ref{finalsection} and use the standard Gaussian trick discussed in section \ref{exacttrans}, to write
\be
\label{newfullint}
Z=\biggl( {N^4\over 2\pi^2}\biggr)^{p} \int   \prod_{i=1}^p dg_i \, d\bar g_i  \, d\mu_i \, d\bar \mu_i \exp N^2 S_{\rm eff}
\ee
where
\be
\label{neweffaction}
\begin{aligned}
S_{\rm eff}=&-\sum_{j=1}^p a_j \mu_j \bar \mu_j +{\rm i}\sum_j (\mu_j \bar g_j + \bar \mu_j g_j) + 
\sum_{\vec k, \vec k'} \alpha_{\vec k , \vec k'} (-{\rm i})^{|\vec k| + |\vec k'|} \Upsilon_{\vec k} (\bar \mu) \Upsilon_{\vec k'}(\mu) \\
&+ F(g_k + b_k, \bar g_k + \bar b_k).
\end{aligned}
\ee
We now write $g_k$ as 
\be
\label{newgrealim}
g_l={1\over 2 l}(\beta_l - i \gamma_l)
\ee
and we also write  
\be
\label{newbrealim}
b_k={1\over 2 k}(\tilde{g}_k - i \hat \gamma_k)
\ee
Performing change of the variables in the integral,
\be
 g_k \rightarrow  g_k + b_k, \quad \bar g_k \rightarrow \bar g_k + \bar b_k.
\ee
 we get,
\be
\begin{aligned}
S_{\rm eff} = &\sum_{j=1}^p \biggl( -a_j \mu_j \bar \mu_j +{{\rm i}\over 2j}  \Bigl( (\beta_j -\tilde{g}_j) (\mu_j + \bar \mu_j)   + {\rm i} (\gamma_j -\hat \gamma_j)(\mu_j - \bar \mu_j) \Bigr) \biggr)
\\
&+\sum_{\vec k, \vec k'} \alpha_{\vec k , \vec k'} (-{\rm i})^{|\vec k| + |\vec k'|} \Upsilon_{\vec k} (\bar \mu) \Upsilon_{\vec k'}(\mu) + F( \beta_k,  \gamma_j).
\end{aligned}
\ee
We will assume that we are analyzing the theory in the ungapped phase, in the proximity of the even multicritical point $m=2k$. In this case we have,
\be
N^2F(\beta, \gamma) = N^2 F_{\rm ug} (\beta, \gamma) + N^2 F_{\rm scaling} (\beta, \gamma),
\ee
where $F_{\rm ug}(\beta, \gamma)$ is the planar free energy in the ungapped phase (\ref{Nexpansion}), and $F_{\rm scaling} (\beta, \gamma)$ 
satisfies
\be
\lim_{N \to \infty}  N^2F_{\rm scaling} (\beta, \gamma) =F^{(m)} (t_{l}),
\ee
where $F^{(m)} (t_{l})$ is the double-scaled free energy at the $m$-th multicritical point determined by the solution to the mKdV hierarchy \cite{Periwal:1990qb}.

To find the saddle point at large $N$ we only have to consider the contribution of the free energy $F(\beta,\gamma)$ in the ungapped phase. The equations for the saddle point are given by,
\be
\label{newsaddleqs}
\begin{aligned}
{\partial S_{\rm eff} \over \partial \beta_j}=& { {\rm i} \over 2 j}(\mu_j +\bar \mu_j) +  {1\over 2j} \beta_j=0, \\
{\partial S_{\rm eff} \over \partial \gamma_j}=&-{ 1 \over 2 j}(\mu_j -\bar \mu_j) + {1\over 2j}  \gamma_j=0 , \\
{\partial S_{\rm eff} \over \partial \mu_j}=&-a_j \bar \mu_j  +{{\rm i} \over 2j} ( \beta_j -\tilde{g}_j+ {\rm i} \gamma_j -{\rm i} \hat \gamma_j) \\ &+ 
 \sum_{\vec k, \vec k'} \alpha_{\vec k , \vec k'} (-{\rm i})^{|\vec k|+ |\vec k'|} {k_j' \over \mu_j} \Upsilon_{\vec k} (\bar \mu) \Upsilon_{\vec k'} (\mu)=0 \\
{\partial S_{\rm eff} \over \partial \bar\mu_j}=& -a_j \mu_j  +{{\rm i} \over 2j} ( \beta_j -\tilde{g}_j - {\rm i} \gamma_j +{\rm i} \hat \gamma_j ) \\ & +
 \sum_{\vec k, \vec k'} \alpha_{\vec k , \vec k'} (-{\rm i})^{|\vec k|+ |\vec k'|} {k_j \over \bar \mu_j} \Upsilon_{\vec k}  (\bar \mu) \Upsilon_{\vec k'} (\mu)=0 \\
\end{aligned}
\ee
In the first two equations we have used that, in the ungapped phase, 
\be
{\partial F_{\rm ug} \over \partial \beta_j}= {1\over 2j} \beta_j, \quad {\partial F_{\rm ug}\over\partial \gamma_j}= {1\over 2j} \gamma_j,
\ee
We will assume that there is a solution to these equations corresponding to the $m$-multicritical even point of the model (\ref{neweffaction}), which 
is characterized by 
\be
\gamma_j=0, \quad \beta_j=\beta_j^{(m)},
\ee
where the critical values of the couplings  $\beta_j^{(m)}$ can be read from the particular solution (\ref{newps}). We find that this solution leads to the conditions
\be
\label{newsolv}
\mu^{(m)}_j=\bar\mu^{(m)}_j={{\rm i}\over 2} \beta^{(m)}_j.
\ee
One finds the equations for the critical submanifolds in the original couplings, $a_j$, $\tilde{g}_k$, and $\alpha_{\vec k, \vec k'}$,
\be
\label{newcriticalplane}
 \beta_j^{(m)} (j a_j -1) + {\tilde{g}^c_j \over j} + \sum_{\vec k, \vec k'}  2^{2 -|\vec k| -|\vec k '|} (-1)^{|\vec k| +|\vec k '|} 
\alpha_{\vec k , \vec k'} {k_j \over \beta_j^{(m)}} \Upsilon_{\vec k +\vec k'}(\beta_j^{(m)})=0, 
\quad j=1, \cdots, p.
\ee
where $\tilde{g}^c_j$ is the critical value of $\tilde{g}_j$, and we have set $\hat \gamma_j^c=0$ for simplicity. 

We now expand the effective action around the critical point, and we expand simultaneously the original couplings $a_j$, $\tilde{g}_j$, $\hat \gamma_j$ and $\alpha_{\vec k, \vec k'}$ around a point $a_j^c$, $\tilde{g}_j^c$, $\hat \gamma_j^c=0$, and $\alpha_{\vec k, \vec k'}^c$ on the critical submanifold determined by (\ref{newcriticalplane}). 
We denote
\be
P(\mu, \bar \mu, \alpha)=\sum_{\vec k, \vec k'} \alpha_{\vec k , \vec k'} (-{\rm i})^{|\vec k| + |\vec k'|} \Upsilon_{\vec k} (\bar \mu) \Upsilon_{\vec k'}(\mu).
\ee
We introduce the column vectors of variables,
\be
\label{newvarias}
\begin{aligned}
\xi(N) n = & \left( \begin{matrix}  \mu_j -\mu_j^{(m)}\\
                                  \bar\mu_j -\bar\mu_j^{(m)} \end{matrix} \right), \quad
\alpha= \left( \begin{matrix} a_j -a_j^c\\
                                  \alpha_{\vec k, \vec k'} -\alpha^c_{\vec k, \vec k'}  \end{matrix} \right), 
                             \\
                             g= &\left( \begin{matrix} \beta_j -\beta_j^{(m)}\\
                                  \gamma_j \end{matrix} \right) , \quad 
b= \left( \begin{matrix} \tilde{g}_j -\tilde{g}_j^c\\
                                 \hat \gamma_j \end{matrix} \right) ,                                   
\end{aligned}
\ee
where $\xi (N)$ is an appropriate scaling factor. When we expand the action in (\ref{neweffaction}) around the $m$-th multicritical point, we obtain
\be
\sum_l \Bigl( g_l {\rm tr}\,  U^l + {\overline g}_l  {\rm tr}\, U^{\dagger l} \Bigr)=V^{(m)} + \sum_{n} N^{n-2m \over 2m+1} t_n \tilde V_n,
\ee
where $V^{(m)}$ is the critical potential associated to the $m$--th multicritical point, and ${\tilde V}_n$ are scaling operators which can be 
explicitly written by using the results of \cite{cdm}. In this way we find the relation between the variables $g$ introduced in (\ref{newvarias}) and the 
scaling operators of the multicritical model,
\be
\label{newrelcrit}
g_a=\sum_{n\ge 0} {\cal G}_{an}N^{n-2m\over 2m+1} t_n,
\ee
where ${\cal G}$ is a matrix that can be explicitly determined from the expressions for the perturbations of the density of eigenvalues. The equation (\ref{newrelcrit}) determines the scaling properties of the $g_a$. 
Notice that we can use the freeedom to rotate $U$ to get rid of one of the $2p$ parameters $g_i, \bar g_i$, 
so we will only have $2p-1$ times. 

We now do a Gaussian integration over $n$. The relevant part of the action 
reads, 
\be
N^2\, S_{\rm eff} = -{1\over 2} N^2 \xi(N)^2 n^t\, {\cal L} \, n + N^2 \xi(N) n^t({\cal J} g -{\cal J} b+ {\cal H} \alpha) + \cdots, 
\ee
where the matrices ${\cal L}$, ${\cal J}$, ${\cal H}$ are given by
\be
\begin{aligned}
{\cal L}= &\left( \begin{matrix}  -{\partial ^2 P \over \partial \mu_j \partial \mu_k} & a_j^{(c)}\delta_{jk} -  {\partial ^2 P \over \partial \mu_j \partial \bar \mu_k}\\
                               a_j^{(c)}\delta_{jk} -  {\partial ^2 P \over \partial \mu_j \partial \bar \mu_k}   &   -{\partial ^2 P \over \partial \bar \mu_j \partial \bar \mu_k} \end{matrix} \right), \\ 
{\cal H}=  &\left( \begin{matrix}  - \bar \mu_j \delta_{jk} & {\partial ^2 P \over \partial \mu_j \partial \alpha_{\vec k, \vec k'}}\\
                               -\mu_j \delta_{jk}   & {\partial ^2 P \over \partial \bar \mu_j \partial \alpha_{\vec k, \vec k'}} \end{matrix} \right), \\ 
 {\cal J}=& {1\over 2} \left( \begin{matrix}  {\rm i} {\cal F} & {\cal F} \\
                               {\rm i} {\cal F}   & -{\cal F} \end{matrix} \right),
                               \end{aligned}
                               \ee
and we have introduced the diagonal matrix 
\be
{\cal F}_{jk}={1\over j} \delta_{jk}, \quad j,k=1, \cdots, p.
\ee
All quantities involved in these matrices are evaluated at the critical point. The Gaussian integration leads to 
\be
\label{newexpg}
N^{2p} ({\rm det} ({\cal L}))^{-{1\over 2}} \exp \biggl\{ {1\over 2} N^2 (g- {\cal E} b-{\cal C} \alpha)^t {\cal M} (g-{\cal E} b- {\cal C} \alpha) + F^{(m)}(t_{\ell}) + \cdots\biggr\},
\ee
where we have assumed that ${\cal L}$ does not have zero modes, and the fact that the Gaussian integration gives an overall factor $N^{-2p}$ which combines with the 
overall $N^{4p}$ in (\ref{newfullint}). Notice that the scaling $\xi(N)$ does not appear in this equation. The choice of $\xi(N)$ must be done in such a way that 
the rest of the terms involving $n$ in the expansion of $N^2 S_{\rm eff}$ vanish in the limit $N\rightarrow \infty$.
The matrices appearing here can be easily obtained from the above data. Then, we have 
\be
\begin{aligned}
{\cal D}=&{1\over 2} \left( \begin{matrix}  {\cal F} & 0 \\
                                 0 & {\cal F}  \end{matrix} \right), \\
{\cal M}=&{\cal J}^t {\cal L}^{-1} {\cal J} +{\cal D} ,\\ 
{\cal C}=&-{\cal M}^{-1} {\cal J}^t {\cal L}^{-1} {\cal H},\\
{\cal E}=&{\cal M}^{-1} {\cal J}^t {\cal L}^{-1} {\cal J}.
\end{aligned}
\ee
Notice that the Hessian associated to $S_{\rm eff}$ is given by
\be
H=  \left( \begin{matrix} - {\cal L} & {\cal J} \\
                                 {\cal J}  & {\cal D}  \end{matrix} \right).
\ee
We now introduce scaling variables for the couplings $g$, $\alpha$.  The scaling of $g$ is determined 
In this way we obtain for (\ref{newexpg}) 
\be
\label{newexpgg}
\exp \biggl\{ {1\over 2} \sum_{n,p} N^{2 + n+p \over 2m+1} (t_n -t^0_n) {\cal A}_{np} (t_p-t^0_p) + F^{(m)}(t_{\ell}) + \cdots\biggr\},
\ee
where 
\be
\label{newscalings}
\begin{aligned}
{\cal A}=&{\cal G}^t {\cal M} {\cal G},\\
t^0_n =&N^{2 m-n\over 2m+1} \sum_{\ell}\Bigl( ({\cal G}^{-1}{\cal C})_{n \ell} {\alpha}_{\ell} +({\cal G}^{-1}{\cal E})_{n j}b_j \Bigr) .
\end{aligned}
\ee
As we see, the scaling of the original coupling constants packaged in $\alpha$, $b$ is determined by the 
scaling of the couplings in the $m$-th critical point. 

In the limit $N\rightarrow \infty$, the integral localizes in 
\be
t_n=t_n^0.
\ee
To see this in detail, we use the following fact. Let $B_{\epsilon}$ be an $n \times n$ matrix whose entries go to $+ \infty$ as 
$\epsilon \rightarrow 0$. Then, one has the following 
\be 
\lim_{\epsilon \to 0}\,  ({\rm det}(B_{\epsilon}))^{1\over 2} e^{-{1\over 2} x^t B_{\epsilon} x} =\pi^{n\over2} \, \delta (x) .
\ee
In our case we find that
\be
\label{newlimitdelta}
\exp \biggl\{ {1\over 2} \sum_{n,p} N^{2 + n+p \over 2m+1} (t_n -t^0_n) {\cal A}_{np} (t_p-t^0_p)\biggr\} \rightarrow
N^{-{\sum_{n\ge 0}(n+1) \over 2m+1}} {\pi^{p-{1\over 2} } \over {\rm det}({\cal G}) ({\rm det}(-{\cal M}))^{1\over2} } \delta(t-t_0)
\ee
as $N\rightarrow \infty$. Remember that there are only $2p-1$ times involved. After changing variables in the integral from $g, \bar g$ to $t$, we inherit a Jacobian 
\be
\label{newjacob}
N^{{\sum_{n\ge 0}(n-2m) \over 2m+1}} {\rm det}({\cal G}).
\ee
Putting all these ingredients together, we finally obtain
\be
Z \sim N ({\rm det}(H))^{-{1\over 2}}\exp F^{(m)}(t_{n}^0),
\label{newfinalresult}
\ee
up to factors of $\pi$. We have assumed here that $H$ has no zero modes. The factor of $N$ comes from the fact that the quotient between the factors of $N$ 
in (\ref{newlimitdelta}) and (\ref{newjacob}) gives a power of $N$ given simply by minus the number of times involved, which is $-2p+1$. This combines with the factor $N^{2p}$ in 
(\ref{newexpg}) to give an overall factor of $N$. In the above derivation we have assumed that ${\cal M}$ (and therefore $H$ has no zero eigenvalues).

We can also analyze the more general case in which ${\cal M}$ (which is a $p\times p$ matrix) has $\ell$ nonzero eigenvalues $d_n$, $n=1, \cdots, \ell$, and 
$2p-\ell$ zero eigenvalues. Let $R^{-1}$ be the orthogonal $2p\times 2p$ matrix that diagonalizes $\cal M$, i.e. $R^{-1t}{\cal M} R^{-1}={\rm diag}(d_n, 0)$. 
Define now the following eigenvectors of ${\cal M}$ 
\be
r =N^{2m \over 2m +1} R g, 
\ee
which in terms of the scaling operators means
\be
r_n =\sum_{q} {\cal R}_{nq} t_q N^{ q \over 2m+1},
\ee
where ${\cal R}=R{\cal G}$. 
Then, the exponent in the Gaussian (\ref{newexpg}) becomes
\be
\label{newzeroexp}
{1\over 2} N^{2 \over 2m+1} \sum_{n=1}^{\ell} d_n \Bigl(r_n -N^{2m \over 2m+1} c_n\Bigr)^2 + N^{2 + 2m \over 2m+1} \sum_{n=\ell+1}^{2p} r_n \zeta_n, 
\ee
where
\be
\begin{aligned}
\zeta_n=&\sum_{q} R^{-1t}_{nq}\bigl({\cal J}^t {\cal L}^{-1} {\cal H} \alpha - {\cal J}^t {\cal L}^{-1} {\cal J}b\bigr)_q,  \quad n =\ell+1, \cdots, 2p\\
c_n=& -d_n^{-1} \sum_{q} R^{-1t}_{nq}\bigl({\cal J}^t {\cal L}^{-1} {\cal H} \alpha - {\cal J}^t {\cal L}^{-1} {\cal J}b\bigr)_q, \quad n=1, \cdots, \ell.
\end{aligned}
\ee
 As $N\rightarrow \infty$, the first term in (\ref{newzeroexp}) gives a delta function constraint of the form
 \be
 \label{newconst}
 \sum_{q\ge 0}{\cal R}_{nq} t_q N^{q\over 2m+1} =c_n, \quad n=1,\cdots,\ell,
 \ee
 therefore there are only $2p-1-{\ell}$ independent times involved. From the behavior of the above equation as $N \rightarrow \infty$ it follows that we have to 
 solve for the times with the higher scaling dimension in terms of the constants $c_n$. This in turn determines the scaling properties of $c_n$:
 \be
 t_q=t_q^0\equiv N^{2m-q \over 2m+1} \sum_{n=1}^{\ell} {\cal R}^{-1}_{qn} c_n, \quad q=2p-1-\ell, \cdots, 2p-2,
\ee
where we have inverted the $\ell \times \ell$ submatrix ${\cal R}_{qn}$, $q,n=2p-1-\ell, p-2$.
This fixes the values of $\ell$ times in the free energy as functions of the scaled parameters $c_n$, $n=1, \cdots, \ell$. 
The other times lead to a integral transform. To see this, let us define
\be
\bar t_q= N^{2m +2+q\over 2m+1}\sum_{n=\ell+1}^{2p} {\cal R}_{nq} \zeta_n .
\ee
This equation determines the scaling of $\zeta_n$. Notice that the scaling properties induced on $c_n$ and $\zeta_n$ are very different. 
Up to overall factors, we end up with the integral
\be
\begin{aligned}
&\int \prod_{n=0}^{2p-2} dt_n \prod_{q=2p-1-\ell}^{2p-2} \delta (t_q-t_q^0) \exp \biggl\{ \sum_{q=0}^{2p-2} t_q {\bar t}_q + F^{(m)}(t_q) \biggr\} =\\
& e^{ \sum_{ q=2p-1-\ell}^{2p-2} t^0_q {\bar t}_q }  \int \prod_{n=0}^{2p-2-\ell}dt_n \exp \biggl\{ \sum_{q=0}^{2p-2-\ell} t_q {\bar t_q}+ F^{(m)}(t_0, \cdots, t_{2p-2-\ell}, t_{2p-1-\ell}^0, \cdots, t_{2p-2}^0) \biggr\} . 
\end{aligned}
\ee
For Hermitian matrix models, a similar result was obtained in \cite{kh}. Notice that the integral transform will change the critical exponents of the model, as noted in \cite{kh}.
\par
To illustrate our formalism we can look on to the example of  free YM theories at finite temperature \cite{Aharony:2003sx,Sundborg:1999ue}
\be
S(U,U^{\dagger})= \sum_{j=1}^{\infty} a_j {\tr}\, U^j \, {\tr}\, U^{\dagger j},
\ee
where
\be
a_j={1\over j}(z_B(x^j) + (-1)^{j+1} z_F(x^j)).
\ee
The equation for the critical surface reduces to 
\be
 \beta_j^{(m)} (j a_j -1) + {\tilde{g}^c_j \over j} =0,
 \ee
and by tuning the value of $\tilde{g}^c_j$ we can reach any critical point. Notice that, if we do not include the $b_k$ terms in the 
original action, only the first critical point $m=1$ can be realized in the model. In that case, one 
has
\be
a_1(T) =1, 
\ee
which defines the Hagedorn temperature $T=T_H$. Also, if we do not 
include the source terms involving $b_k$, we can turn on only a single scaling operator in the 
theory and we recover the $m=1$ model. When one includes the $b_k, \bar b_k$ couplings one can also recover all the evolution times of the double-scaled matrix 
model.


\section{Acknowledgment}
We would like to thank Ofer Aharony, Jose Barbon, Avinash Dhar, Rajesh Gopakumar, David Gross, Veronica Hubeny, Elias Kiritsis, Gautam Mandal, Hong Liu, Mukund Rangamani and Ashoke Sen, for useful discussions during the course of this work. We are especially thankful to Shiraz Minwalla for many valuable and critical discussions during the course of this work. PB and SRW would like to thank the CERN theory division for a visit where part of this work was done. Various versions of this work were presented at various string meetings: Third Regional Crete Meeting in String theory (Crete, June 2005), IIT Kanpur String theory workshop (IIT Kanpur, October 2005), Inauguration Meeting of the Institute for Quantum Spacetime (Seoul, October 2005), Batsheva de Rothschild Seminar on Innovative Aspects of String Theory, India-Israel string theory meeting (Ein-Boqeq, March 2006), 12th Regional Conference on Mathematical Physics (Islamabad, March 2006), Workshop on Blackholes, Blackrings and Topological String Theory (Munich, April 2006), IPM String School and Workshop (Tehran, April 2006). We thank the organizers of these conferences for the invitation to present this work.

\end{document}